\documentclass[twocolumn,dvipsnames]{aastex62}
\usepackage{amsmath,amstext}
\usepackage{apjfonts}
\usepackage[figure,figure*]{hypcap}
\usepackage{ulem}
\usepackage{xspace}
\usepackage{xcolor}
\usepackage{lineno}
\usepackage{xfrac}
\usepackage{graphicx}
\usepackage{natbib}
\usepackage{multimedia}
\usepackage{amssymb}
\usepackage{amsmath}
\usepackage{array,booktabs}
\usepackage{mathptmx}
\usepackage{booktabs}
\usepackage{hyperref}
\usepackage{float}
\usepackage{todonotes}
\usepackage{multirow}

\newcommand{\erg}{\,{{\rm erg}}}

\newcommand{\msun}{\,{M_{\odot}}}

\newcommand{\s}{\,{{\rm s}}}

\newcommand{\cm}{\,{{\rm cm}}}

\shorttitle{The Landscape of Collapsar Outflows}

\shortauthors{Ore Gottlieb}

\begin{document}

\title{The Landscape of Collapsar Outflows:\\Structure, Signatures and Origins of Einstein Probe Relativistic Supernova Transients}

\author[0000-0003-3115-2456]{Ore Gottlieb}
\email{ogottlieb@flatironinstitute.org}
\affiliation{Center for Computational Astrophysics, Flatiron Institute, 162 5th Avenue, New York, NY 10010, USA}
\affil{Department of Physics and Columbia Astrophysics Laboratory, Columbia University, Pupin Hall, New York, NY 10027, USA}
\affil{Department of Physics and Kavli Institute for Astrophysics and Space Research, Massachusetts Institute of Technology, Cambridge, MA 02139, USA}

\begin{abstract}

The Einstein Probe is revolutionizing time-domain astrophysics through the discovery of new classes of X-ray transients associated with Type Ic-Broad Line supernovae. These events commonly exhibit bright early-time optical counterparts, and sudden afterglow rebrightening within the first week -- features that existing models fail to explain. In particular, structured jet and cocoon scenarios are inconsistent with the observed sharp rebrightening and multi-day optical emission, while the refreshed shock model is ruled out due to its inconsistency with collapsar hydrodynamics. Drawing on 3D general-relativistic magnetohydrodynamic simulations, we present the multi-scale angular and radial structure characterizing collapsar outflows. The resulting morphology features episodic, wobbling jets with a ``top-hat'' geometry, embedded within a smoother global cocoon and disk ejecta angular structure. The wobbling jets give rise to variations in radiative efficiency that can account for the observed alternation between X-ray-dominated and $\gamma$-ray-dominated jet emission. The top-hat structure of individual wobbling jet episodes naturally explains the sudden rebrightening observed when the emission from the top-hat jet cores enters the observer's line of sight. The radial structure is consistent with that inferred from observations of stripped-envelope supernovae. It comprises a mildly relativistic cocoon ($0.3\lesssim\beta\Gamma\lesssim3$) that may power an early ($\sim1\,{\rm day}$) rapidly decaying emission, followed by slower, black hole accretion disk-driven outflows ($\beta\lesssim0.3$), which dominate the slowly evolving optical emission at $t\gtrsim1\,{\rm day}$. This novel multi-component outflow structure provides a unified explanation for the multiband light curves observed in Einstein Probe transients and is likely a common feature of Type Ic-Broad Line supernovae more broadly.

\end{abstract}

\section{Introduction}\label{introduction}

Recent years have seen a major progress in the discovery and characterization of fast, luminous astrophysical transients across the electromagnetic spectrum, driven primarily by the synergy between wide-field surveys and rapid-response follow-up observations \citep[e.g.,][]{Prentice2018,Ho2019,Ho2020,Ho2022,Margutti2019,Perley2019,Perley2021,Coppejans2020,Bright2021,Yao2021,Sun2022,Sun2025,Chen2023,Matthews2023,Nicholl2023,Gutierrez2024,Migliori2024}. A major milestone in the field was the launch of the Einstein Probe (EP) in January 2024, which has transformed transient astronomy. Equipped with its Wide-field X-ray Telescope (WXT), EP has not only uncovered a new population of high-redshift fast X-ray transients, but its precise localization capabilities have facilitated comprehensive, systematic multi-wavelength campaigns, which are revolutionizing our understanding of the broad spectrum of fast transient phenomena.

Several EP-identified fast X-ray transients have clear associations with $\gamma$-ray bursts \citep[GRBs;][]{Levan2024,Liu2025} and/or broad-lined Type Ic supernovae \citep[SNe Ic-BL;][]{Aryan2025,EylesFerris2025,Izzo2025,Rastinejad2025,Srinivasaragavan2025,Sun2025,vanDalen2025}, pointing at the collapse of rapidly rotating massive stars \citep[``collapsars'';][]{Woosley1993,Galama1998,Woosley2006} as a prominent source of EP transients. These events span a broad range of dynamical phenomena, from early-time direct GRB emission, through sharp non-thermal multiwave rebrightening, to evolving optical emission from sub- and mildly relativistic outflows.

One natural candidate for producing the sub- and mildly relativistic hot plasma is the ``cocoon,'' generated during the jet's propagation within the collapsing star \citep{Meszaros2001,RamirezRuiz2002,Matzner2003,Lazzati2005}. Cocoon emission has been predicted to produce vigorous UV/optical signals \citep{Nakar2017}, potentially associated with SNe Ib/c \citep{Piran2019}, and is hypothesized as a possible source of fast optical transients \citep{Gottlieb2022b}. Indeed, cocoon emission is a popular interpretation for several puzzling EP transients, including EP240414a \citep{Hamidani2025,vanDalen2025,Zheng2025}, EP240801a \citep{Jiang2025}, EP241021a \citep{Gianfagna2025}, and EP250108a \citep{EylesFerris2025,Yadav2025}\footnote{Although \citet{Yadav2025} argue against quasi-spherical mildly relativistic cocoons, we note that mildly relativistic cocoons must inherently exhibit some degree of collimation. In fact, the cocoon structure naturally arising during jet propagation constitutes the ``structured jet'' proposed by \citet{Jiang2025,Yadav2025} \citep[see][for further discussion on the cocoon-jet interplay]{Gottlieb2021a}.}.

Although cocoon emission arises naturally in collapsar scenarios, observations indicate that the observational signatures of EP events are more complex than those predicted by cocoon models. The cocoon represents only one component of a richer system involving multiple interacting outflows. This includes a proto-magnetar (PM), expected to form prior to black hole (BH) collapse \citep{Duncan1992}, and launch its own outflows \citep[e.g.,][]{Wheeler2000,Uzdensky2007,Bucciantini2009,Metzger2011,Metzger2015}, as well as a BH accretion disk that powers magnetically-driven subrelativistic ejecta \citep{Bopp2025}. In addition, the first 3D collapsar simulations have shown that collapsar jets possess not only angular, but also azimuthal and radial structure \citep{Gottlieb2022c}, offering an opportunity to infer jet morphology from the irregular emission signatures observed in EP events.

In this \emph{Letter}, we revisit several of the most intriguing EP transients through the lens of collapsar physics, aiming to connect specific emission features with their underlying physics. We argue that both wobbling jets and disk ejecta likely play critical roles in shaping the observed rebrightening and thermal optical emission, respectively. The structure of the paper is as follows: in \S\ref{sec:collapsar}, we review the evolution of collapsars and outline the structure of the resulting outflows. In \S\ref{sec:emission}, we argue that current theoretical models fail to account for the EP observations and compare the electromagnetic signatures arising from the various components of collapsars to EP observations. We summarize our results and conclude in \S\ref{sec:conclusions}.

\section{Physics of Collapsars}\label{sec:collapsar}

The collapse of rapidly rotating stars leads to the formation of a spinning PM \citep[e.g.,][]{Raynaud2020,Masada2022,White2022}, which may eventually collapse into a spinning BH. Such a BH can launch relativistic jets via the Blandford-Znajek \citep[BZ;][]{Blandford&Znajek1977} mechanism. Sustained jet launching capable of powering GRBs through the BZ mechanism \citep{Popham1999,MacFadyen&Woosley1999} typically requires three conditions \citep[e.g.,][]{Symbalisty1984, Metzger2008, Gottlieb2022a}: (i) a dimensionless BH spin parameter $a \gtrsim 0.3$ \citep{Gottlieb2023}; (ii) the presence of a steady accretion disk surrounding the BH; and (iii) sufficient poloidal magnetic flux available to the BH. \citet{Gottlieb2024b} demonstrated that an accretion disk may form around the PM before its collapse; such a disk can inhibit magnetic reconnection, enabling the BH to inherit the PM's magnetic flux and thus relaxing the third requirement. Consequently, an accretion disk around a moderately spinning BH emerges as the only prerequisite for BZ jet launching. However, the angular momentum profiles of stellar progenitors and the likelihood of disk formation during the PM phase remain uncertain \citep[see e.g.,][]{Fryer2025}.

\begin{figure*}[]
  \centering
  \includegraphics[width=0.9\textwidth,trim={0cm 0cm 0cm 0cm}]{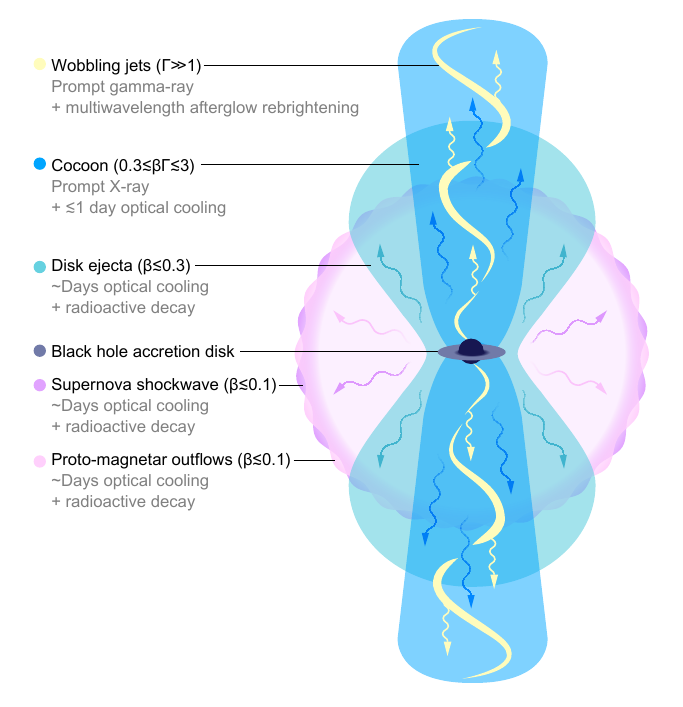}
  \caption{
  Schematic illustration of collapsar outflows and emission:\\
  \emph{Structure}:
The collapse results in the formation of a proto-magnetar, which launches subrelativistic outflows (violet) prior to collapsing into a black hole. The ensuing formation of a black hole surrounded by an accretion disk drives moderately collimated, magnetized disk ejecta (turquoise) and a wobbling ultra-relativistic episodic jet powered by the Blandford-Znajek mechanism (yellow). As the jet propagates, it shocks the surrounding stellar material, generating a collimated, mildly relativistic cocoon (blue). An additional subrelativistic SN shockwave from core bounce (magenta) is possible; however, the proto-magnetar and/or disk ejecta are consistent with the energy and velocity distributions observed in stripped-envelope SNe, rendering the SN shockwave unnecessary to account for observations. We note that the illustrated collimation of the cocoon and disk ejecta represents characteristic angles where the bulk of the energy is concentrated. In practice, the outflows mix in a spherical morphology, with the energy distributed as a function of angle.\\
  \emph{Emission}:
Observers detect intermittent emission as the jets aligned with their line of sight generate $ \gamma$-ray emission, while misaligned jets and cocoon produce lower-efficiency X-ray emission. Both on-axis and off-axis observers may detect a sharp afterglow rebrightening when the emission from a ``top-hat'', wobbling jet episode enters their line of sight. During the first day, the thermal optical cooling emission is dominated by the cocoon, after which the disk ejecta, and later the proto-magnetar outflows, take over. Enhanced nucleosynthesis in the disk and proto-magnetar outflows promotes $^{56}{\rm Ni} $ mixing in the outer regions of the ejecta, producing the bright $^{56}{\rm Ni}$-powered peak in the light curve.
}
  \label{fig:sketch}
\end{figure*}

Following the collapsar evolution discussed above, collapsar events are expected to produce at least four types of outflows: (i) PM outflows; (ii) accretion disk ejecta; (iii) BZ jets; and (iv) cocoon. In this section, we examine the magnetohydrodynamic processes that give rise to this morphology and describe the resulting structure of each outflow component. Figure~\ref{fig:sketch} presents an overview of the resulting structure and the corresponding emission. We discuss the latter in detail in \S\ref{sec:emission}.

\subsection{Types of Outflows}

\subsubsection{Proto-magnetar outflows}

A few numerical simulations have explored the formation and evolution of PM outflows in collapsing stars \citep[e.g.,][]{Mosta2014,Mosta2015,Mosta2018,Kuroda2020,Aloy&Obergaulinger2021,Obergaulinger&Aloy2022,Shankar2025}. These studies find that, although PM outflows carry substantial energy ($E \gtrsim 10^{51}\,\erg$), they generally exhibit only mild collimation and low velocities ($\beta \lesssim 0.1$), primarily due to the intense ram pressure exerted by the infalling stellar material. As a result, Fig.~\ref{fig:sketch} illustrates that the energy deposited by PM outflows may contribute as a power source similar to a quasi-spherical SN explosion. However, if PM outflows delay the PM collapse for several seconds, long enough for outflows to clear its immediate surroundings, subsequent PM-driven outflows can achieve higher velocities of $\beta \lesssim 0.5$ \citep{Aloy&Obergaulinger2021}. Analytic estimates based on typical magnetic field strengths and mass accretion rates suggest that the PM generally collapses within $ \sim 1\,\s $ into a moderately spinning BH, with a spin parameter of $a \sim 0.3$ \citep{Gottlieb2024b}. This rapid collapse implies that PM outflows may not persist long enough to unbind the stellar envelope or accelerate to mildly relativistic velocities.

\subsubsection{Disk Ejecta}

Except for magnetar-driven outflows, all other collapsar outflows require the presence of an accretion disk. Hydrodynamic axisymmetric collapsar simulations suggest that collapsar disks can drive powerful hydrodynamic outflows \citep{Dean2024a,Dean2024b,Fujibayashi2024,Menegazzi2024,Menegazzi2025}. However, 3D neutrino-general-relativistic magnetohydrodynamic simulations indicate that such outflows are unable to overcome the ram pressure of the infalling gas \citep[e.g.,][]{Issa2025}. In contrast, collapsar disks threaded by dynamically important magnetic fields, as expected around BHs that launch jets, power their own magnetically driven sub-relativistic outflows \citep{Bopp2025,Issa2025} via the Blandford-Payne mechanism \citep{Blandford&Payne1982} and through magnetic flux eruptions \citep[e.g.,][]{Tchekhovskoy2011,Chatterjee2022,Ripperda2022}. Recent simulations by \citet{Bopp2025} demonstrated that magnetized collapsar disks produce disk ejecta with isotropic-equivalent energy $ E_{\rm iso} \approx 10^{52}\,\erg $, breaking out of the stellar envelope at subrelativistic velocities.

\subsubsection{Wobbling Jets and Cocoon}\label{sec:jets}

\begin{figure*}[]
  \centering
  \includegraphics[width=0.522\textwidth]{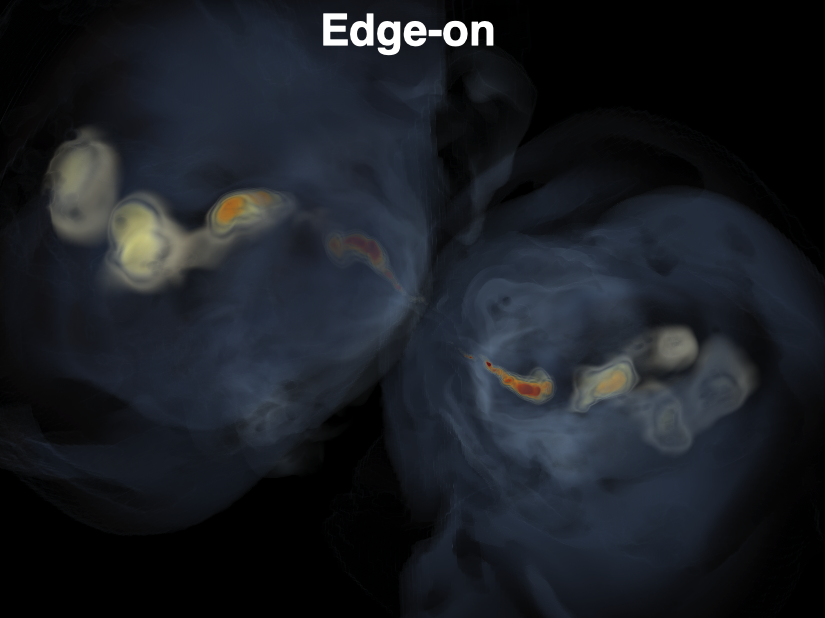}
  \includegraphics[width=0.444\textwidth]{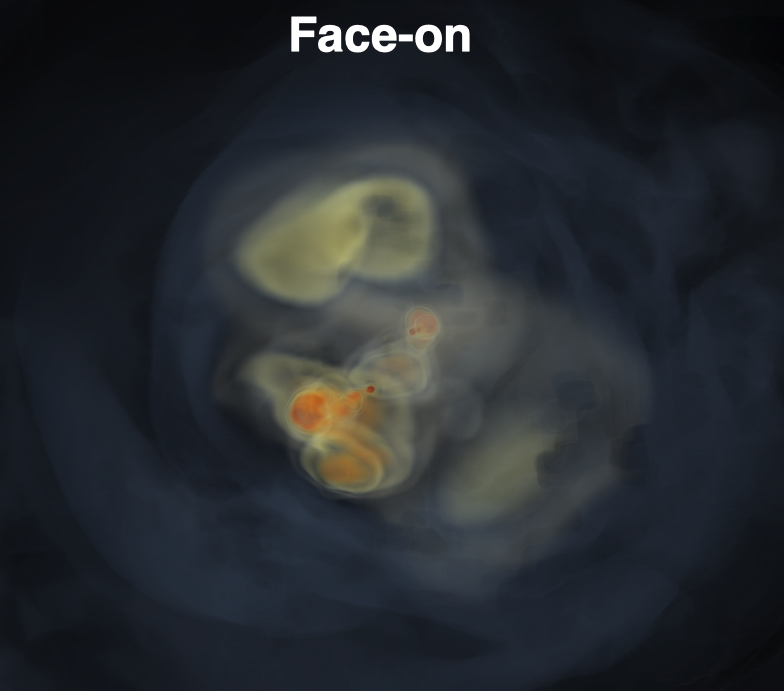}
  \caption{
    3D renderings of a collapsar jet simulation from \citet{Gottlieb2022c} illustrating the velocity of relativistic outflows when they reach 10 stellar radii, viewed edge-on (left) and face-on (right). Relativistic jet elements appear in red, slower components in yellow, and the surrounding cocoon in blue. The left panel demonstrates the wobbling and episodic nature of collapsar jets. During phases when none of the jet episodes is directed toward the observer, the cocoon and the misaligned off-axis wobbling jets may produce low-efficiency softer emission. This leads to intermittency and spectral variability in the GRB prompt emission along a given line of sight (e.g., at $ \theta = 0 $ on the right panel). As the outflow decelerates during the afterglow phase, the light curve initially decays, followed by rising off-axis wobbling jet emission as misaligned relativistic jet components gradually enter the observer's beaming cone. This results in potentially multiple rebrightening episodes in the afterglow light curve.
  }
  \label{fig:Jet}
\end{figure*}

If the BH possesses a moderate spin of $a \gtrsim 0.2$, a sufficient amount of magnetic flux will activate the BZ mechanism, launching bipolar relativistic jets. The jets interact with the dense stellar core, producing a shocked cocoon composed of mildly relativistic jet plasma and spherical non-relativistic shocked stellar material \citep{Bromberg2011b}. As the cocoon expands, it envelops and interacts with the accretion disk. Large-scale, long-duration, high-resolution 3D general-relativistic magnetohydrodynamic collapsar simulations by \citet{Gottlieb2022c} followed the post-BH formation phase, and demonstrated the self-consistent formation of both the accretion disk and jets from the collapsing star and through the BZ mechanism. These simulations revealed that jet–disk interactions transfer stochastic momentum to the disk, destabilizing it and continuously altering its orientation. Consequently, the jet, which is anchored to the disk's angular momentum, develops a wobbling motion at its base.\footnote{Although only a handful of global collapsar simulations exist due to their high computational cost, we have consistently observed wobbling in all of our global collapsar runs, including direct-horizon collapsars \citep{Gottlieb2025}, with additional setups in preparation. The underlying driver of this behavior is the inevitably strong disk--cocoon interaction. By contrast, in post–compact-object mergers, the disk--cocoon interaction is considerably weaker because there is no sustained infalling envelope as in collapsars, leading to correspondingly weaker wobbling \citep[e.g.,][]{Gottlieb2022d,Gottlieb2023c}. This emerging trend aligns with the expected dependence of wobbling on the strength of the disk--cocoon interaction. Nevertheless, further numerical work is required to establish that this behavior is robust to currently missing physics, such as neutrino-cooled disks in global simulations.} Figure~\ref{fig:Jet}, taken from \citet{Gottlieb2022c}'s 3D simulations, demonstrates that this motion is imprinted onto the jet's larger-scale structure and remains observable even after the jet emerges from the collapsing star on a breakout time of a few seconds.

The duration over which the BH and disk launch jets and outflows depends on how long they can maintain accretion and high magnetic flux. In collapsars, the accretion disk mass is continuously replenished by stellar material falling inward, sustaining a steady mass accretion rate. Thus, the duration of collapsar-driven GRBs is expected to align with the characteristic accretion timescale \citep{Matzner2003}. For a typical Wolf-Rayet progenitor star with radius $R \sim R_\odot$, the density begins to decline steeply at $ r \gtrsim 10^{10}\,\cm $, corresponding to a free-fall time of $ \sim 100\,\s $, comparable to the typical collapsar GRB duration \citep{Kouveliotou1993,McBreen1994}.

\subsection{Outflow Structure}\label{sec:overall_structure}

While the relative importance of each outflow component depends on the specific stellar properties, certain general relationships can be robustly drawn. PM outflows, carrying energies of $E \sim 10^{51}-10^{52}\,\erg$, are emitted early and briefly during the collapse. As a result, they are subject to substantial mixing, significantly slowing their expansion, transforming them into a quasi-spherical explosion, which may hinder their ability to produce distinct observational signatures. Due to the lack of large-scale numerical simulations tracing PM outflows beyond the stellar envelope, their exact contribution to the final outflow structure remains uncertain.

The relative contribution between the jet-cocoon structure and disk ejecta depends on the BH spin. Provided sufficient magnetic flux is supplied to the BH, either by the PM \citep{Gottlieb2024b} or through dynamo processes \citep{Shibata2025}, two regimes are relevant:\footnote{The jet power depends on the BH spin and the magnetic flux. Since collapsar jet launching requires a high magnetic flux \citep{Issa2025}, this imposes a constraint on the BH spin, as high spins would produce jets with luminosities exceeding those observed in GRBs. As a result, spins of $a \gtrsim 0.5$ are ruled out \citep{Gottlieb2023}.} moderately spinning BHs ($0.2 \lesssim a \lesssim 0.5$) that launch GRB jets; and slowly spinning BHs ($a \lesssim 0.2$) that do not produce jets capable of powering GRBs. At $  a\approx 0.25$, the power of the disk ejecta is comparable to that of the BZ jets, $ L_{\rm disk} \approx L_{\rm BZ} $ \citep{McKinney2012, Narayan2012, Tchekhovskoy2012, Bopp2025}. Given that jets require a persistent accretion disk, the disk and jet launching durations are also expected to be comparable, leading to similar total energies, $E_{\rm disk} \approx E_{\rm BZ}$. Since jets deposit about half of their energy into the cocoon \citep{Bromberg2011b, Gottlieb2021a}, we expect $E_{\rm disk} \approx 2E_{\rm cocoon} \approx 2E_{\rm jet} $. Conversely, in the case of slowly spinning BHs, the disk ejecta becomes the dominant energy channel, $E_{\rm disk} \gg E_{\rm jet} $.

\begin{figure}[]
  \centering
  \includegraphics[width=0.5\textwidth]{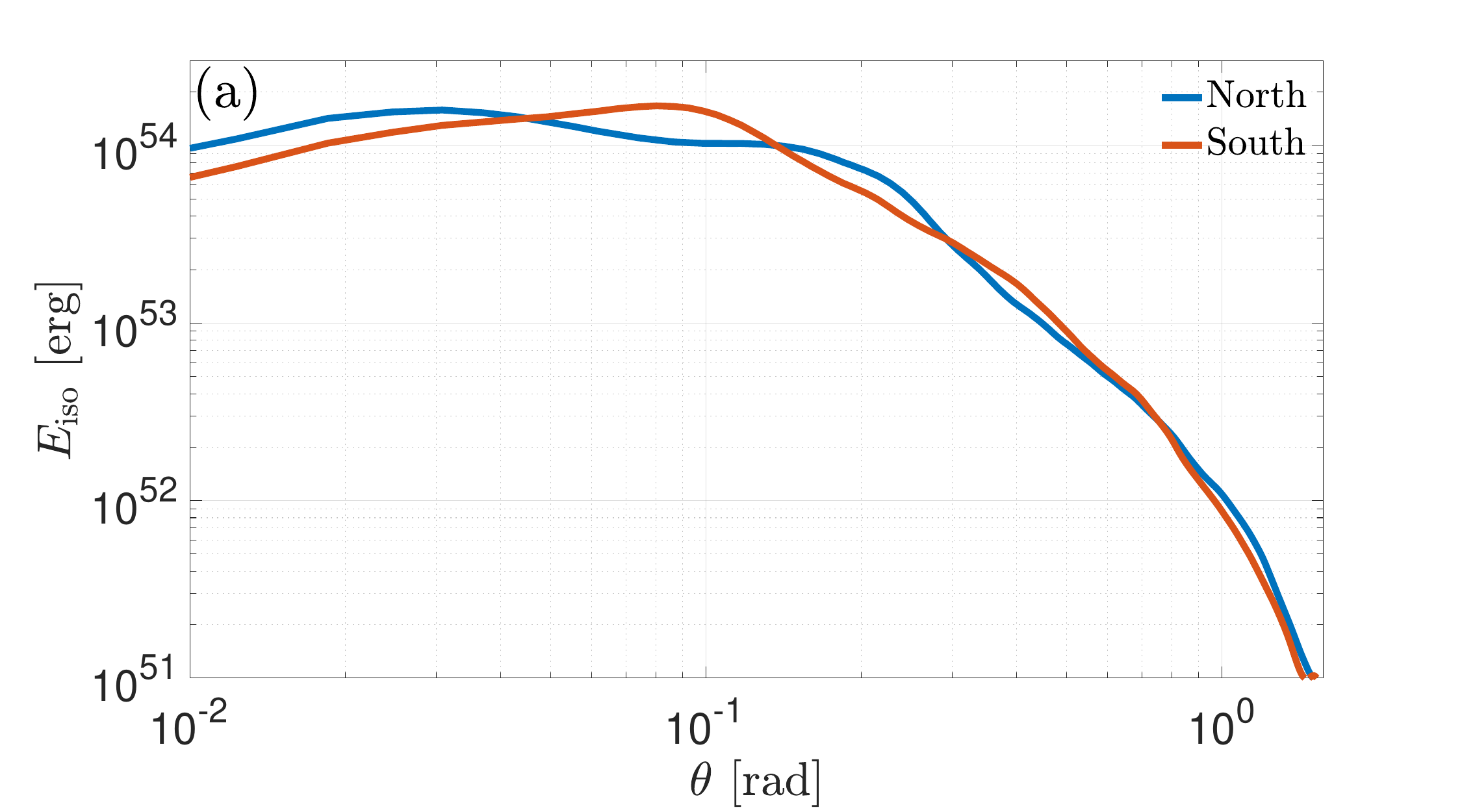}
  \includegraphics[width=0.5\textwidth]{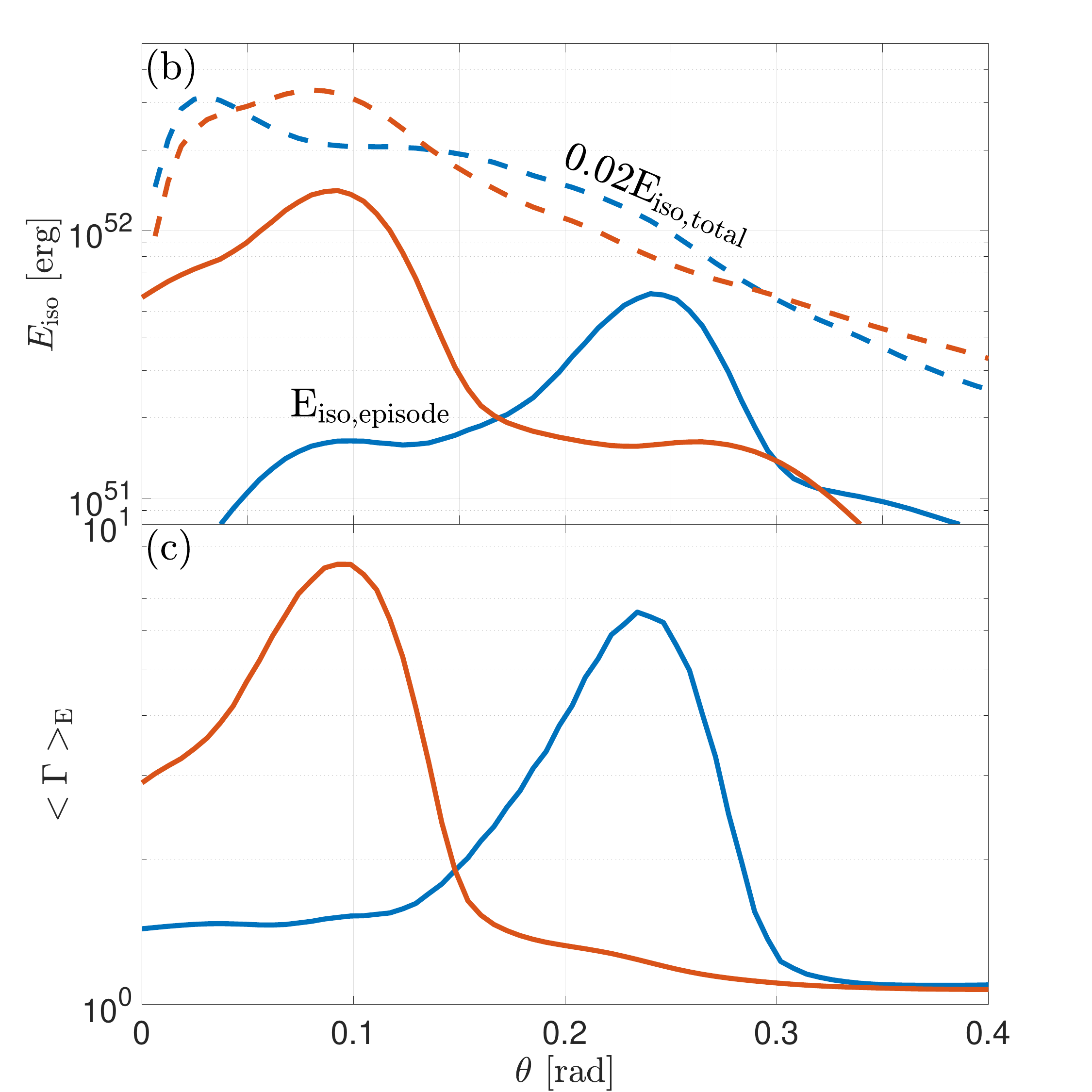}
  \caption{{\bf (a)}: Radially integrated (north hemisphere in blue and south hemisphere in red) isotropic-equivalent energy, as a function of angular separation from the stellar axis of rotation ($ \theta = 0 $). The global outflow shows a gradual decline in energy with angle. {\bf (b)}: Isotropic equivalent energy and {\bf (c)}: Energy density-weighted Lorentz factor through a shell containing a bipolar wobbling jet (jet and its counter-jet rotated by $ 180^\circ $) episode at $ \sim 3 $ stellar radii (delineated in dark red in Fig.~\ref{fig:Jet}). The two peaks shown by the solid lines indicate that the jets' cores deviate from the rotation axis due to wobbling. Both jets exhibit steep angular gradients (top-hat) in Lorentz factor and isotropic-equivalent energy, in contrast to the smoother global angular profiles obtained for the full volumetric distribution (whose renormalized shape is shown in dashed lines in (b) for comparison).
  }
  \label{fig:angle_dist}
\end{figure}

Ultimately, all outflow components interact and merge into a single structured outflow with both radial and angular profiles. Figure~\ref{fig:angle_dist}(a) presents the radially integrated isotropic-equivalent energy as a function of angular separation from the stellar rotation axis. The resulting global angular structure features a flat core with power-law wings \citep{Gottlieb2021a}, and remains unaffected by jet wobbling \citep{Gottlieb2022c}, as it reflects radial and azimuthal integrations over multiple jet episodes launched at varying angles. However, the angular structure of individual jet episodes has not been investigated until now. Figure~\ref{fig:angle_dist}(b,c) depicts $ E_{\rm iso} $ (b) and Lorentz factor (c) across a shell containing a single bipolar wobbling jet episode. Both jets show a prominent deviation from the rotational axis due to the wobbling motion. In stark contrast to the smooth core+power-law structure seen in the global outflow structure (dashed lines in panel b), single jet episodes exhibit a distinct core and steep power-law wings, whose power-law indices exceed $ \sim 5 $ for both the energy and Lorentz factor. Accordingly, the angular profile of single jet episodes is more characteristic of a top-hat jet than a structured jet. We note that the relatively low jet Lorentz factor of $ \Gamma \lesssim 10 $ in 
Fig.~\ref{fig:angle_dist}(c) arises from the floor values imposed in numerical simulations, which constrain the asymptotic Lorentz factor. Nevetheless, \citet{Gottlieb2022c} showed that a higher asymptotic Lorentz factor, as suggested by observations, does not alter the global outflow morphology, and is expected to sharpen the episodic top-hat structure of the relativistic jet relative to the mildly relativistic cocoon, thereby further accentuating its top-hat profile.

Previous numerical studies have consistently shown that efficient jet-star mixing produces a complex radial profile in the jet, where more heavily entrained material moves at lower velocities, whereas more baryon-free jet components retain relativistic speeds \citep{Gottlieb2019b,Gottlieb2021a}. The jet-cocoon interaction redistributes energy uniformly across the logarithm of the velocity space between the escape velocity from the star and the jet Lorentz factor ($ \beta_{\rm esc} \lesssim \beta\Gamma \lesssim \Gamma_{\rm jet}$), consistent with energy equipartition between the jet and the cocoon. This result appears robust and universal, having been consistently reproduced in simulations of hydrodynamic \citep{Gottlieb2021a}, magnetized \citep{Gottlieb2022c}, and choked \citep{Eisenberg2022} jets. However, observationally, \citet{Piran2019} inferred that stripped-envelope SNe exhibit excess energy in non-relativistic ejecta. Consequently, \citet{Eisenberg2022} concluded that the jet-cocoon alone cannot fully account for the observed velocity distributions, suggesting the necessity of an additional channel supplying non-relativistic ejecta. The non-relativistic ejecta contains $ E_{\rm Ic-BL} \sim 10^{52}\,\erg $ \citep[e.g.,][]{Cano2013,Taddia2015,Taddia2019,Prentice2016}, about an order of magnitude more energy than a typical GRB jet energy\footnote{Assuming typical beaming angle and radiative efficiency from \citet{Beniamini2016}.}, $ E_{\rm jet} \approx E_{\rm cocoon} \sim 10^{51.5}\,\erg $ \citep{Minaev2020}.

\begin{figure}[]
  \centering
  \includegraphics[width=0.5\textwidth]{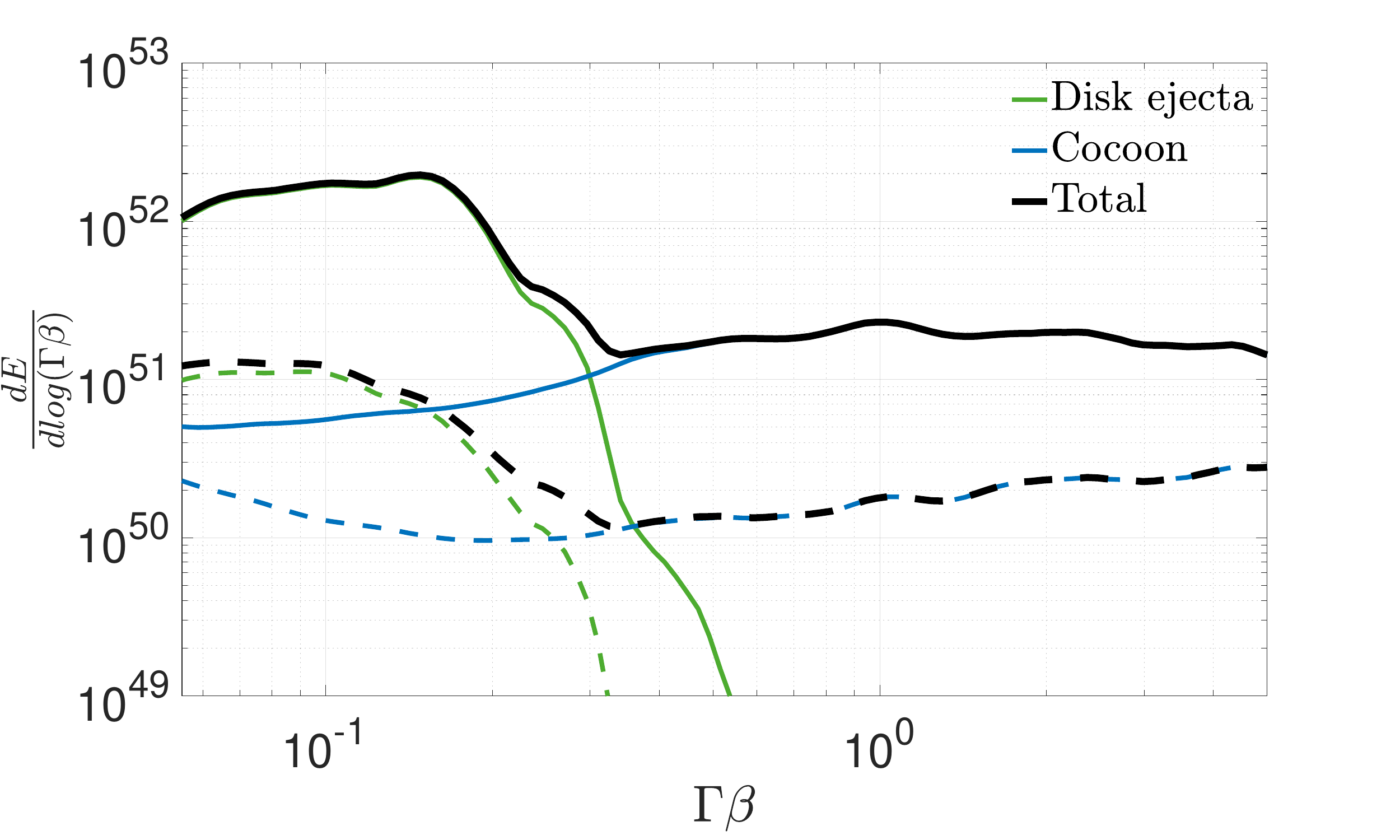}
  \caption{Total energy distribution per logarithmic interval of asymptotic proper velocity of collapsar outflows. Solid lines represent the total energy distribution for disk ejecta (green), cocoon (blue), and their combined total (black). The disk ejecta dominates the energy at $ \beta \lesssim 0.3 $, whereas the cocoon exhibits a flat distribution and dominates at mildly relativistic velocities, consistent with the velocity distribution inferred from stripped-envelope SNe. Dashed lines indicate the thermal energy fraction of the respective total energies. The distribution is truncated at low velocities ($\beta_{\rm min} \approx 0.05$), as material moving at slower velocities has not escaped the stellar envelope by the end of the simulations.}
  \label{fig:Egb}
\end{figure}

We estimate the energy distribution of collapsar outflows by combining numerical results of disk ejecta from non-spinning BHs \citep{Bopp2025}, which isolate the disk ejecta contribution from jet-cocoon interference, with jet-cocoon structures from rapidly spinning BHs, where the jet-cocoon component dominates. Figure~\ref{fig:Egb} depicts the resulting energy distribution per logarithmic interval of $\Gamma\beta$, normalized to characteristic values of $E_{\rm disk} = 10^{52}\,\erg$ and $E_{\rm cocoon} = 10^{51.5}\,\erg$, typical of SNe Ic-BL. The distribution reveals two components: non-relativistic disk ejecta bump at $ \beta \lesssim 0.2 $, and a flat distribution corresponding to mildly relativistic cocoon material. This distribution matches the one inferred from observations of stripped-envelope SNe \citep[see][]{Piran2019}. This result was also anticipated by \citet{Hayakawa2018}, who estimated that if the disk ejecta is continuously launched and moderately collimated within the star, before transitioning to spherical expansion after breakout, it can reproduce the observed emission properties of GRB-associated SNe.

\section{Emission}\label{sec:emission}

Building on the outflow structure described in \S\ref{sec:collapsar}, we examine the expected emission signatures associated with each component and connect them to the origins of selected collapsar-associated EP transients. We divide the discussion into three observational signatures: GRBs, non-thermal afterglow rebrightening, and optical counterparts.

\subsection{GRB Emission}\label{sec:GRB}

The WXT has so far detected three X-ray transients coinciding with GRBs: EP240219a \citep{Yin2024} with GRB 240219A detected by Fermi/GBM; EP240315a \citep{Levan2024,Liu2025} with GRB 240315C detected by Swift/BAT \citep{DeLaunay2024} and Konus-Wind \citep{Svinkin2024}; and EP240801a \citep{Jiang2025} with GRB 240801A detected by Fermi/GBM. Only GRBs 240315C and 240801A were followed by detectable GRB afterglow emission. All three events exhibit X-ray emission lasting considerably longer than the corresponding GRBs, with durations of $ \sim 10^2-10^3\,{\rm s} $. The observed prolonged soft X-ray emission is likely a consequence of the high redshift associated with these events. After redshift correction, the rest-frame durations appear consistent with the expected free-fall timescales of collapsar progenitors.

In these GRBs, the timing of the $\gamma$-ray detection relative to the overall X-ray light curve varies: GRB 240219A aligned with the onset of the X-ray signal, while in the other events, the $\gamma$-ray emission occurred at an intermediate phase during the X-ray emission. This non-monotonic spectral evolution may arise from a radially mixed jet structure, where variations in the baryon loading, Lorentz factor, internal energy, and magnetization along the jet produce alternating episodes of harder/brighter and softer/dimmer emission \citep[e.g.,][]{Ito2015,Gottlieb2019b}. Alternatively, as shown in Fig.~\ref{fig:Jet}, the light curve evolution may reflect the complex three-dimensional structure of wobbling jets, in which harder emission appears when the relativistic jet epoch (red in Fig.~\ref{fig:Jet}) aligns with the observer's line of sight, and softer emission is observed when off-axis jet episodes dominate the emission. Both the radial structure and the latitude-longitude jet structure can produce recurring hard–soft transitions in the observed emission. The fact that only a single $\gamma$-ray bright epoch is observed in each of the three events supports the expectation of a complex jet angular (latitude-longitude) structure, rather than a simple jet with purely radial variation. Events in which the X-ray emission lacks a $ \gamma $-ray counterpart may arise from off-axis jet or cocoon emission. Such cases can occur either when the viewing angle lies just outside the wobble angle, or within the wobble angle but outside the beaming cones of all wobbling jet episodes.

\subsection{Afterglow Rebrightening}\label{sec:afterglow}

Multiwavelength observations in the days following the X-ray detection reveal compelling evidence for synchrotron self-absorbed afterglow emission. This is observed both in events accompanied by a GRB, such as EP240315a \citep{Gillanders2024,Ricci2025}, and in those without a detected GRB, such as EP240414a \citep{Sun2025} and EP241021a \citep{Busmann2025}. In all cases, the multi-wavelength emission indicates an ongoing relativistic jet activity. The jets drive a blast wave that expands into the surrounding circumstellar medium (CSM), where it decelerates as it sweeps up mass that is energetically comparable to the initial kinetic energy of the ejecta. This deceleration leads to the continuous fading behavior observed in GRB afterglows. However, in EP240315a, EP240414a, and EP241021a, a rebrightening has been observed in the optical bands (and also in X-rays in EP240315a). This behavior challenges the classical GRB afterglow scenario, raising two fundamental questions about the rebrightening: (i) Why is it detected only in some events? and (ii) What is its physical origin?

\subsubsection{Challenges of Proposed Models}

The detection of GRBs, along with the direct association of some events with SNe Ic-BL, supports a collapsar origin and thus motivates a focus on collapsar models rather than other events such as accreting white dwarfs or binaries \citep{Wu2025}, or tidal disruption events \citep{Xinwen2025}, whose afterglows evolve over considerably longer timescales \citep[e.g.,][]{Alexander2020}. As density fluctuations in the CSM cannot account for the observed sharp features \citep{Nakar2007}, a few alternative models have been proposed to explain the afterglow rebrightening. Two widely discussed models proposed to explain the rebrightening are off-axis emission from a jet with angular structure \citep[e.g.,][]{Yadav2025,Zheng2025}, and refreshed shocks \citep[e.g.,][]{Busmann2025,Ricci2025,Srivastav2025}, or a combination of both \citep{Gianfagna2025}. However, both of these models face significant challenges.

Structured jets can naturally explain the absence of $\gamma$-ray emission when the observer's line of sight initially misses the jet core, and the rebrightening emerges as the jet core gradually enters the line of sight. However, the very steep rise observed in EP241021a \citep[$ L_\nu \sim t^4$;][]{Busmann2025} and EP240414a \citep[$ L_\nu \sim t^3$;][]{Zheng2025}, cannot be reproduced by a smooth angular jet structure \citep{Lamb2021}, unless the jet is composed of discrete segments \citep{Gianfagna2025} with distinct Lorentz factors \citep{Xinwen2025}, similar to top-hat jets \citep{Granot2002,Nakar2002}. Such a segmented configuration of two-component relativistic jets \citep[see][]{Granot2005,Beniamini2020,Li2023}, previously proposed to explain rebrightening features \citep{Liang2013}, naturally arises from wobbling top-hat jets, as we demonstrated in \S\ref{sec:overall_structure}, and as we argue in \S\ref{sec:wobbling_rebrightening}, likely accounts for the observed sharp rebrightening behavior. In contrast, the assumed structures in \citet{Gianfagna2025,Yadav2025,Zheng2025} of jets with only an angular profile formed through continuous interaction with the ambient medium, will produce a monotonic jet+cocoon structure \citep[see jet structure profiles in][]{Gottlieb2021a}, leading to a gradual and prolonged rise, similar to that in GW170817 \citep[e.g.,][]{Mooley2018}. Similar challenges arise for radial structures, which are characteristic of choked jets (see \S\ref{sec:240414}).

An alternative model that has gained recent popularity is the refreshed shock scenario \citep{Panaitescu1998,Rees1998,Sari2000,Granot2003}. In this model, the rebrightening is attributed to a late injection of slower moving material with Lorentz factor $\Gamma_{\rm s} $, which eventually catches up with the decelerating forward shock initially moving at $\Gamma_{\rm f} > \Gamma_{\rm s} $, once the latter slows down to $\Gamma < \Gamma_{\rm s} $. Although this model has been appealing for explaining sudden rises in afterglow light curves, it faces several critical challenges, which we outline below.

First, the Lorentz factor of the emitted shell is given by $\Gamma_{\rm sh} = L_{\rm sh}/\dot{m}_{\rm sh} c^2$, where $\dot{m}_{\rm sh} $ is the baryon entrainment rate in the shell, and $ L_{\rm sh} $ is the shell luminosity. In order to produce a noticeable rebrightening, most of the jet energy needs to be injected with the slower shell at later times, $ L_{\rm s} > L_{\rm f} $ \citep[e.g.,][]{Rees1998,Granot2003}. This implies that the baryon entrainment must increase substantially over time, $ \dot{m}_{\rm s} > \dot{m}_{\rm f} $, to compensate for the decreasing Lorentz factor associated with the increasing shell luminosity. However, jet elements launched early are subject to intense mixing, which gradually subsides as the jet carves a channel through the dense stellar interior. Consequently, later-emitted jet elements experience considerably less mixing and can maintain considerably higher Lorentz factors \citep{Gottlieb2019b,Gottlieb2021a}. This evolution is in direct tension with the requirements of the refreshed shock model.

Second, for refreshed shocks to be efficient, they require two shells to collide at a radius of $R \sim 10^{15}-10^{16}\,\cm \approx 10^{10}$ BH gravitational radii. Expecting a relativistic shell to remain coherent over $10^{10}$ dynamical timescales is highly unrealistic. In practice, the shell's structure will gradually smooth out due to continuous interaction with the surrounding outflow \citep[see e.g.,][]{Gottlieb2020c}. As a result, any resulting shock would likely be weak if it forms at all, as its energy would already be redistributed. These considerations suggest that the refreshed shock model is overly simplistic, highly fine-tuned, and unlikely to operate in realistic GRB environments.

Finally, even if such a refreshed shock were to occur, it would likely be inconsistent with the EP observations. In EP241021a, attributing the steep rise to a refreshed shock requires that it follows a jet break, yet no jet break was observed prior to the rebrightening \citep{Gianfagna2025}. The isotropic-equivalent prompt energy in $\gamma$-rays and X-rays of the jet in EP240315a is $E_{\rm iso} \sim 10^{54}\,\erg $ \citep{Levan2024}, implying that the refreshed shock would need to be extremely energetic yet only mildly relativistic to produce a noticeable rebrightening. This requirement further underscores the implausibility of the refreshed shock scenario in explaining these events.

\subsubsection{Wobbling Jets}\label{sec:wobbling_rebrightening}

As discussed in \S\ref{sec:jets}, collapsar simulations show that jets are inherently non-axisymmetric and develop a complex three-dimensional structure involving angular, azimuthal, and radial components. This naturally results in non-monotonic energy and velocity distributions with distinct emission zones. Such structure has several observational consequences. The combination of jet afterglow and the absence of $\gamma$-ray detection (when not attributed to redshifted $ \gamma $-rays) in some EP events, such as EP241021a, can be naturally explained if the observer does not detect on-axis emission from the wobbling jets. Instead, the observed signal arises from softer, weaker emission from off-axis jet episodes and/or slower shocked jet material in between the relativistic wobbling jet elements, producing the required lower radiative efficiency of $\epsilon_\gamma \lesssim 1\%$ \citep{Yadav2025}. Once one of the energetic jet cores sufficiently decelerates to enter the observer's line of sight, the top-hat-like structure of the wobbling jets embedded within the global cocoon structure (Fig.~\ref{fig:angle_dist}) predicts a sudden rise in the light curve. Recurrent rebrightening may also occur if additional energetic jet segments at larger angular distances and/or at initially higher Lorentz factors subsequently enter the observer's view \citep[e.g.,][]{Granot2005,Li2023}. In contrast, in axisymmetric jets where both energy and Lorentz factor decrease monotonically with angle, deceleration leads to lateral spreading and a gradual leakage of energy into the observer's line of sight, resulting in a slow, single, smooth rise in the light curve \citep[see e.g.,][]{Beniamini2020}.

The wobbling jet structure found in collapsar simulations indicates a characteristic jet opening angle of $ \theta_{\rm jet} \approx 0.1\,{\rm rad} $ and a wobbling angle of $ \theta_w \approx 0.3\,{\rm rad} $ \citep{Gottlieb2022c}. Taking this structure as fiducial, we compare it with values inferred from rebrightening times. Assuming a constant density CSM profile \citep{Schulze2011}, we consider the time evolution of the blast wave Lorentz factor \citep[e.g.,][]{Sari1998,Razzaque2010},
\begin{equation}\label{eq:Gamma}
    \Gamma(t) = 6\,(1+z)^{3/8}\left(\frac{E_{k, \rm iso}}{10^{53}\,\erg}\frac{1\,\cm^{-3}}{n}\right)^{1/8}\left(\frac{t}{\rm day}\right)^{-3/8},
\end{equation}
where $z$ is the redshift, $E_{k, \rm iso}$ is the isotropic-equivalent kinetic energy of the blast wave, and $n$ is the CSM number density. 

The high-redshift ($z = 4.9$) event EP240315a exhibited a rebrightening at $t_{\rm rb} \approx 13\,{\rm hr}$. Assuming the kinetic energy is comparable to the high-energy emission, $E_{k, \rm iso} \approx E_{\rm iso} \approx 10^{54}\,\erg$, and $n = 1\,\cm^{-3}$, we find from Equation~\eqref{eq:Gamma} that $\Gamma \approx 20$, implying the observer was located $\Delta\theta \approx 0.05\,{\rm rad}$ off-axis from the most energetic segment of the wobbling jet. The jet afterglow in EP240414a at $z = 0.4$ indicates $E_{k, \rm iso} \approx 10^{51}\,\erg$ with $n = 1\,\cm^{-3}$ for the initially observed decelerating jet prior to brightening \citep{Sun2025}. Assuming a typical \emph{total} (over all episodes) jet isotropic equivalent energy of $E_{k, \rm iso} \approx 10^{53}\,\erg$ at $ t_{\rm rb} \approx 1\,{\rm day} $, we find $ \Gamma \approx 7 $, implying that $\Delta\theta \approx 0.15\,{\rm rad}$. For EP241021a ($z = 0.75$), which exhibited a rebrightening at $ t_{\rm rb} = 6\,{\rm days} $, we adopt $E_{k, \rm iso} = 10^{53}\,\erg$ and $n = 10^{-3}\,\cm^{-3}$ based on afterglow modeling by \citet{Yadav2025}. This yields $\Gamma \approx 9$, corresponding to $\Delta\theta \approx 0.1\,{\rm rad}$. A secondary, weaker rebrightening, which could be powered by a weaker wobbling jet episode, was observed at $t_{\rm rb} = 20.7\,{\rm days}$ \citep{Xinwen2025}, which corresponds to $\Delta\theta \approx 0.17\,{\rm rad}$. We conclude that all inferred $0.05 \lesssim \Delta\theta \lesssim 0.2 $ values are consistent with the fiducial wobbling jet structure, characterized by a wobbling angle of $\theta_w \lesssim 0.3\,{\rm rad}$.

A similar consistency check can be applied to the afterglow peak time. For instance, in EP240414a, at the time of the optical peak $ t_p \approx 4\,{\rm days} $, the spectral regime corresponds to $ \nu_a < \nu_m < \nu_{\rm opt} < \nu_c $ ($ \nu_a $, $ \nu_m $, $ \nu_{\rm opt} $ and $ \nu_c $ denote the self-absorption, synchrotron, optical and cooling frequencies, respectively). The off-axis afterglow calibrated peak time in this regime is \citep{Gottlieb2019a}
\begin{equation}
    t_p = 4\,\left(\frac{E_{\rm jet}}{10^{51}\,\erg}\frac{1\,{\rm cm^{-3}}}{n}\right)^{1/3}\left(\frac{\Delta\theta}{0.1\,{\rm rad}}\right)^2\,{\rm days}\,,
\end{equation}
indicating that the peak timescale is consistent with the physical parameters above.

\subsection{Optical Follow-up}\label{sec:optical}

A few EP events exhibit intriguing optical counterparts, with detections ranging from within a day of the first X-ray light to several weeks post-explosion. These counterparts culminate in the emergence of the SN $ ^{56} $Ni decay-powered peak \citep[e.g.,][]{Rastinejad2025,vanDalen2025}. The origin of the early-time optical emission remains, however, debated. In the following, we explore the physical mechanisms underlying the optical emission in two particularly interesting cases, EP240414a and EP250108a, which display markedly different evolutionary behaviors.

\subsubsection{EP250108a}

When no jet afterglow emission is detected, whether due to a large viewing angle, the absence of a relativistic jet, or an intrinsically weak jet, optical signatures that would otherwise be outshone by the jet can be detected. A notable example is EP250108a, which exhibited a fast X-ray transient \citep{Li2025} followed by a declining optical signal preceding the SN main peak at $ t \approx 12\,{\rm days} $ \citep{EylesFerris2025,Rastinejad2025,Srinivasaragavan2025}. To explain the optical emission prior to the $^{56}{\rm Ni} $ emission, several models have been proposed, including SN ejecta-CSM interaction and cocoon cooling emission \citep{EylesFerris2025,Srinivasaragavan2025,Zhu2025}. However, as we show below, neither model can sufficiently account for the observed week-long decay.

The inferred photospheric velocity declines from $\beta \approx 0.3$ at $t \approx 1$ day after the explosion, to $\beta \approx 0.1$ at $t \approx 5\,{\rm days}$ post-explosion \citep{EylesFerris2025,Srinivasaragavan2025}. This observation alone disfavors a spherical explosion powered solely by SN ejecta, as it is difficult for a spherical shock to reach such high velocities. Instead, it suggests a more collimated outflow, as expected from collapsars.

Cooling emission from the cocoon, accompanied by a jet or a choked jet in the CSM (see Appendix~\ref{sec:pluto}), has been proposed as the origin of the thermal optical component. However, the cocoon properties discussed in \S\ref{sec:collapsar} are inconsistent with several key observational signatures. First, as shown in \S\ref{sec:overall_structure}, the disk ejecta dominates over the cocoon at velocities $ \beta \lesssim 0.3 $, implying that the cocoon may dominate the emission only at $ t \lesssim 1\,{\rm day} $, as the disk ejecta governs the emission at later times. Indeed, the fits for the cocoon emission presented by \citet{EylesFerris2025,Srinivasaragavan2025} suggest a characteristic velocity of $ \beta \sim 0.1 $, as appropriate for the disk ejecta (or PM outflows) rather than the cocoon\footnote{The cocoon may possess additional energy at low velocities if the jet is choked inside the star, allowing sufficient mixing between the cocoon and unshocked stellar material to deposit all the cocoon energy in velocities $ \beta \lesssim 0.3 $. However, for the cocoon to dominate over the disk ejecta, the jet must carry a substantial amount of energy, making it increasingly difficult for it to be choked deep within the star.}.

Second, the universal energy distribution (\S\ref{sec:overall_structure}) has a distinct prediction for the cooling emission shape. Consider an energy distribution, $ E(\beta) \propto \beta^{-\epsilon} $, for a subrelativistic plasma that expands adiabatically, the bolometric luminosity scales as \citep{Piro2018,Gottlieb2022a}
\begin{equation}\label{eq:L}
    L(t_{\rm obs}) \sim t_{\rm obs}^{-\frac{4}{3+\epsilon}}\,,
\end{equation}
where the observed time evolves with the emitting gas velocity as
\begin{equation}\label{eq:time}
    t_{\rm obs} \sim \beta^{-\frac{3+\epsilon}{2}}\,.
\end{equation}
Regardless of whether the jet is choked or not, the cocoon maintains $ \epsilon \approx 0 $ \citep[e.g.,][]{Eisenberg2022}, resulting in cocoon cooling emission that follows $ L \sim t_{\rm obs}^{-4/3} $. Equation~\eqref{eq:time} indicates that the transition from the cocoon-dominated emission at $ \beta \approx 0.5 $ to disk emission at $ \beta \approx 0.3 $ occurs over about a factor of two in time, implying that by $ \sim 2\,{\rm days} $, the disk ejecta dominates the emission. As the emitting region shifts to the slower-moving disk ejecta, the energy distribution shown in Fig.~\ref{fig:Egb} indicates an initial phase where $ \epsilon \gg 1 $, leading to a roughly constant bolometric luminosity $ L \sim {\rm const} $ (Eq.~\eqref{eq:L}). Since the disk ejecta is significantly more massive than the cocoon, the evolution in velocity space slows down (Eq.~\eqref{eq:time}), resulting in a prolonged plateau phase in the light curve.

\begin{figure}[]
  \centering
  \includegraphics[width=0.5\textwidth]{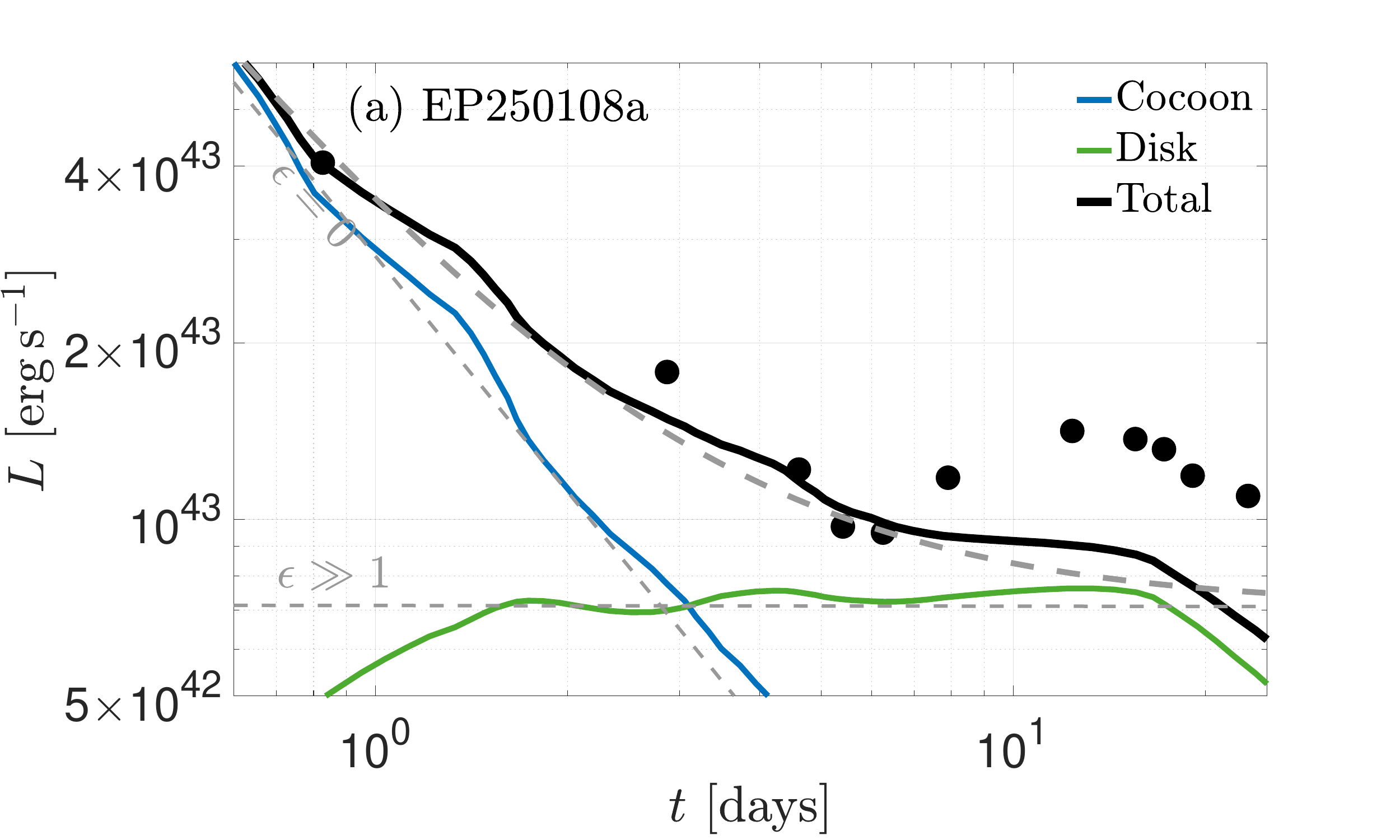}
  \includegraphics[width=0.5\textwidth]{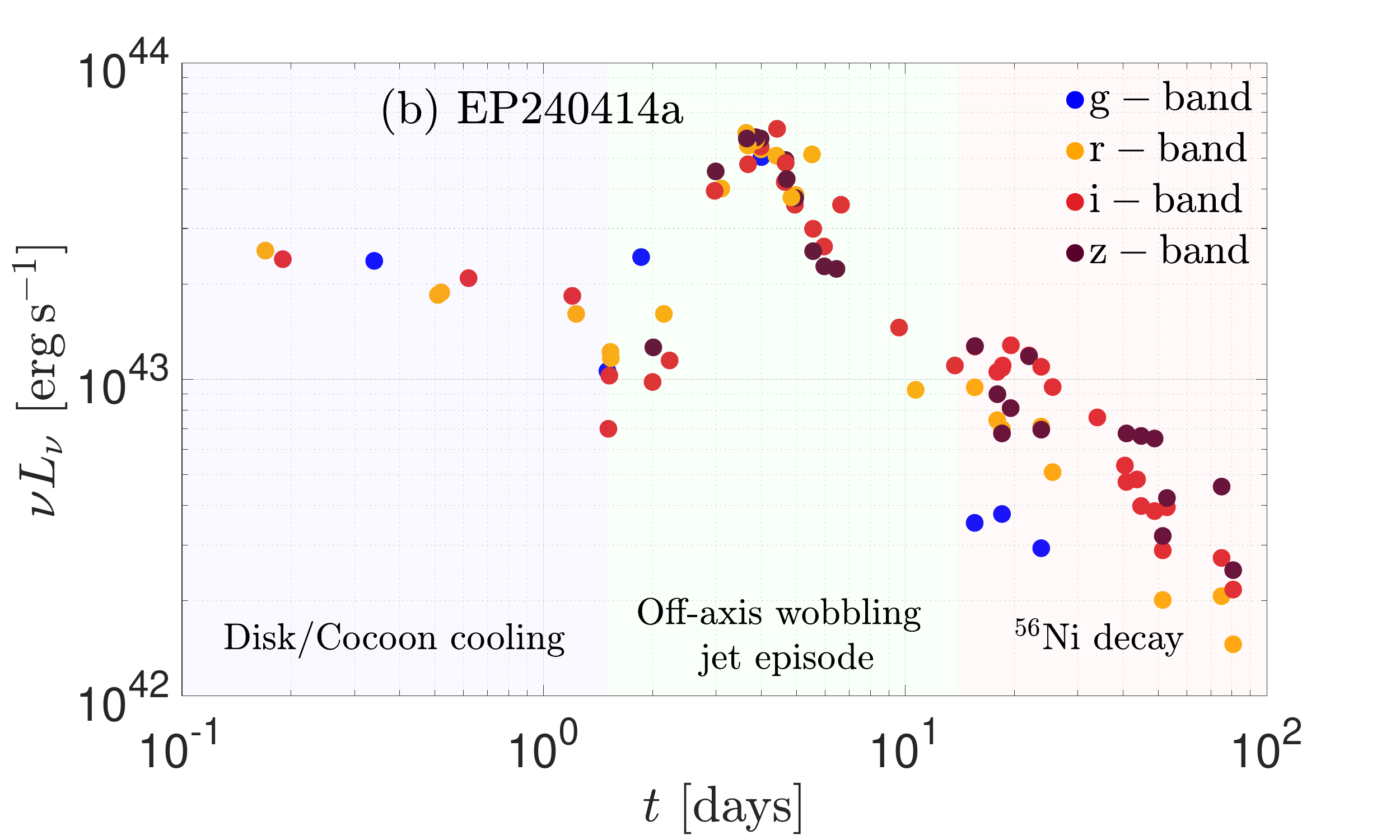}
  \caption{{\bf (a)}: Analytic (dashed gray lines) and numerical (solid lines) calculations of cocoon (blue), disk ejecta (green), and total (black lines) cooling emission are compared with the estimated bolometric luminosity of EP250108a (black dots) \citep{Srinivasaragavan2025}. The temporal evolution indicates that cocoon emission may contribute to the first data point at $t \approx 1\,{\rm day}$, but it decays too rapidly to account for the later observations. The disk ejecta emission dominates the observations at $ 3\,{\rm days} \lesssim t \lesssim 7\,{\rm days}$, before the main SN peak at $ t \approx 12\,{\rm days}$.
  {\bf (b)}: $ \nu L_\nu $ light curve of EP240414a across optical/infrared bands \citep{Srivastav2025,Sun2025,vanDalen2025} presents three phases of optical evolution: cooling emission produces the shallow power-law decline during the first day (blue-shaded background); a subsequent peak powered by a non-thermal afterglow from a wobbling top-hat jet (green-shaded background); and the emergence of thermal emission generated by $ ^{56}{\rm Ni} $-decay (red-shaded background).}
  \label{fig:L}
\end{figure}

Figure~\ref{fig:L}(a) depicts the numerical (solid lines) calculation, based on the thermal energy distribution in Fig.~\ref{fig:Egb}, and the analytic (dashed gray lines) calculation, compared to the bolometric luminosity estimate (black dots) from \citet{Srinivasaragavan2025}. The analytic estimate assumes $ \epsilon = 0 $ for the cocoon and $ \epsilon \gg 1 $ for the disk, demonstrating a good agreement between the two calculations, showing that the cocoon emission (blue) dominates the emission in the first couple of days before the transition to the disk ejecta emission (green).

We conclude that in EP250108a, the observed data are inconsistent with the cocoon cooling emission, which decays too fast, favoring disk ejecta as the dominant emission source during $ 3\,{\rm days} \lesssim t \lesssim 7\,{\rm days} $, prior to the emergence of the main SN signal. While the simulation in \citet{Bopp2025} does not track the gas composition of the disk ejecta, axisymmetric studies have shown that collapsar accretion disks \citep[e.g.,][]{Zenati2020,Dean2024b,Fujibayashi2024,Menegazzi2024} and PM outflows \citep[e.g.,][]{Suwa2015,Wang2016,Moriya2017} can serve as primary production sites of $ ^{56}{\rm Ni} $ in stripped-envelope SNe, yielding significantly larger amounts than those found in ordinary core-collapse SNe. The launching of massive disk ejecta, potentially preceded by PM outflows, enhances mixing efficiency and can account for the substantial $ ^{56}{\rm Ni} $ mass in the outer layers of the ejecta in EP250108a \citep{Rastinejad2025}, as well as in stripped-envelope SNe more broadly \citep[e.g.,][]{Maeda2003,Maeda2003b,Ashall2019,Moriya2020}.

\subsubsection{EP240414a}\label{sec:240414}

The X-ray afterglow of EP240414a was followed by optical and radio observations spanning nearly three orders of magnitude in time, from $ \sim 0.1\,{\rm day} $ to several months post-burst \citep{Srivastav2025,Sun2025,vanDalen2025,Bright2025}. Fig.~\ref{fig:L}(b) presents $ \nu L_\nu $ across multiple optical/infrared bands, featuring three phases. Initially, the emission exhibits a shallow power-law decline during the first $ \sim 1\,{\rm day} $ (pale blue background). This is followed by a pronounced rebrightening that peaks at $ t \approx 4\,{\rm days} $ (pale green background), and subsequently declines before the emergence of $ ^{56}{\rm Ni} $ peak a couple of weeks later (pale red background). At $ t \sim 30\,{\rm days} $, the radio afterglow peaks as the self-absorption frequency becomes comparable to the radio frequencies.

The combination of luminous X-ray emission and red optical colors observed in EP240414a during the first two phases indicates that the emission is non-thermal \citep{Srivastav2025}. While the second phase is consistent with non-thermal afterglow emission, spectroscopic observations during the first day reveal that the optical counterpart's color is inconsistent with a pure afterglow origin \citep{vanDalen2025}. This suggests that the optical and X-ray emission originate from different physical processes: the early X-rays are likely dominated by synchrotron emission, which may also contribute to the optical band, although the optical emission is likely dominated by cooling.

\citet{Hamidani2025} proposed that the emission could originate from a jet that barely escaped the dense CSM with a reduced Lorentz factor. They found that cocoon cooling blended with jet afterglow emission could reproduce the observed luminosity if it possesses a high energy budget of $ E \sim 10^{52}\,\erg $, corresponding to a barely choked jet with $ E_{\rm iso} \sim 10^{54}\,\erg $. From a hydrodynamic standpoint, choking such an energetic jet might be challenging. In Appendix~\ref{sec:pluto}, we present a 3D simulation of a relativistic jet emerging from a star and propagating through a CSM, and demonstrate that choking such a jet would require exceptionally massive and extended CSM. Furthermore, the observed sharp rebrightening is difficult to reconcile with this model, as choked jets are expected to produce a slowly evolving afterglow \citep[see, e.g., the numerical calculations in][]{Nakar2018}, as also illustrated by the comparison between the choked jet model and the observed data in figure 2 of \citet{Hamidani2025}. These arguments disfavor the scenario of an outflow resembling that of a low-luminosity GRB.

If the thermal emission peaks in the optical bands at $ T_{\rm BB} \gtrsim 2\times 10^4\,{\rm K} $ \citep{vanDalen2025}, then the cocoon cooling emission in the observed bands lies in the Rayleigh-Jeans regime. In this case, while the bolometric luminosity of cocoon emission declines as $ L \sim t_{\rm obs}^{-4/3} $, the \emph{spectral} luminosity at the observed bands may feature the observed shallow decay of $ L \sim t_{\rm obs}^{-0.2} $ \citep{Zheng2025}. This suggests that the initial optical phase could be powered by cocoon cooling emission, while the rebrightening phase is primarily due to the jet afterglow. Alternatively, the early emission may originate from subrelativistic disk ejecta, as also suggested by the fit of \citet{Zheng2025}, who inferred a velocity of $\beta \sim 0.1$ for the emitting gas, indicating a disk ejecta or PM outflows origin. In either case, the structured outflows required to explain the optical emission cannot account for the observed sharp rebrightening, which is more consistent with the behavior of a top-hat jet arising from a wobbling jet configuration (see \S\ref{sec:afterglow}). Figure~\ref{fig:angle_dist} illustrates how wobbling jets can give rise to locally top-hat-like jet episodes embedded within a broader, global cocoon structure. This hybrid morphology can simultaneously account for both the early cocoon emission and the pronounced rebrightening observed when the emission from a top-hat jet episode enters the observer's line of sight. 

\section{Conclusions and Discussion}\label{sec:conclusions}

Einstein Probe is transforming the field of time-domain astrophysics, not only by uncovering new X-ray transients but also by providing precise localization that enables detailed follow-up observations. This capability allows for a more comprehensive understanding of the evolution and diversity of transient phenomena, particularly those linked to collapsars. In this \emph{Letter}, we build on state-of-the-art numerical simulations to present a coherent picture of the evolution of the various components in collapsars and to connect their observational signatures with stripped-envelope SNe and the newly discovered Einstein Probe transients.

Figure~\ref{fig:sketch} illustrates that collapsar outflows exhibit a fully three-dimensional structure, characterized by both radial and angular variations. Relativistic jets, which wobble due to interactions between the black hole accretion disk and the surrounding cocoon, produce multiple intermittent episodes at varying angles. As these wobbling jets interact with the collapsing stellar envelope, they deposit a substantial fraction of their energy, inflating a mildly relativistic cocoon. The resulting global angular structure displays a gradual decline with angle in both energy and Lorentz factor. Individual jet episodes, however, retain distinct top-hat profiles embedded within the broader cocoon.

Observations of stripped-envelope SNe indicate the presence of a subrelativistic, energetic component that cannot be accounted for by the cocoon alone. We find that the radial structure of the outflow, comprising a mildly relativistic cocoon and subrelativistic ejecta from the black hole accretion disk, matches well the velocity distribution inferred from stripped-envelope SN observations. The cocoon dominates the outflow at velocities in the range $ 0.3 \lesssim \beta \lesssim 0.95$, and may power the early optical emission through cooling-envelope radiation within the first day. At lower velocities $ \beta \lesssim 0.3 $, the disk-driven shocked ejecta becomes dominant, likely producing the optical signal at $ t \gtrsim 1\,{\rm day} $. At even lower velocities, proto-magnetar outflows may contribute additional cooling emission at later times. Nucleosynthesis in the disk and/or PM outflows enhances the $ ^{56}{\rm Ni} $ abundance in the outer ejecta layers and gives rise to the bright $ ^{56}{\rm Ni} $ decay-powered peak.

Some observational features appear to be common among multiple Einstein Probe transients, including alternating hard and soft prompt emission, a sharp afterglow rebrightening within the first $\sim 1\,{\rm week}$ that cannot be reproduced by either structured or choked jet models, and bright optical counterparts during the first few days post-explosion. The alternating hard $\gamma$-ray and softer X-ray emission components are consistent with on-axis jet elements and off-axis wobbling jet/cocoon emission, respectively. We argue that the refreshed shock scenario is unrealistic as it is inconsistent with the hydrodynamics of collapsars, and therefore cannot account for the sudden afterglow rebrightening. Instead, the observed afterglow rebrightening can be attributed to a wobbling top-hat jet episode viewed off-axis. The optical emission suggests that optical emission transitions from cocoon-dominated during the first $\sim 1\,{\rm day}$, to disk ejecta-dominated afterwards. For scenarios involving jets choked in the circumstellar medium, we show that the required circumstellar medium mass must be considerably higher than predicted by analytic models (see Appendix~\ref{sec:pluto}).

Based on three Einstein Probe events associated with SNe but lacking $\gamma$-ray burst detections: EP250108a, EP240414a, and EP250304a, \citet{Rastinejad2025} estimated the event rate of this transient class to be $\sim10-100\,{\rm Gpc}^{-3}\,{\rm yr}^{-1}$, significantly higher than the $\gamma$-ray burst rate of $\sim 1\,{\rm Gpc}^{-3}\,{\rm yr}^{-1}$ \citep[e.g.,][]{Wanderman2010}. This higher rate is broadly consistent with a simple cocoon emission scenario, where an axisymmetric jet surrounded by a cocoon in which the bulk of the energy is within an opening angle of $\theta_c \approx 0.3$. However, given that $\gamma$-ray burst jets likely wobble, the actual ratio of cocoon-only to jet events is lower. This implies that some events are not powered by cocoon emission, either due to the absence of relativistic outflows (e.g., if the black hole spin is too low) or because the observer is located too far off-axis. In such cases, the optical follow-up may reveal only disk ejecta-driven emission (e.g., in EP250108a), or exclusively SN emission (as in EP250304a) if, e.g., a magnetized disk (and thus jet, cocoon, and disk ejecta) fails to form.

We conclude that while axisymmetric jet and cocoon models have been extensively studied, collapsars produce a far more complex outflow structure. Numerical simulations highlight the roles of wobbling jets, disk ejecta, and proto-magnetar winds as key components in interpreting stripped-envelope SNe and Einstein Probe transients. In particular, the comparable energy expected from proto-magnetar outflows and disk ejecta of $ \sim 10^{52}\,\erg $ is consistent with stripped-envelope SN observations, raising the question of which component dominates the emission. These insights call for detailed numerical and analytic investigations of the dynamics and emission from each outflow component, from the prompt phase through the late afterglow, and how they may be observationally distinguished from canonical axisymmetric jet-cocoon models.

\acknowledgements
I am grateful to Lucy Reading-Ikkanda/Simons Foundation for designing Figure~\ref{fig:sketch}. I thank Hamid Hamidani and Jianhe Zheng for helpful comments, and Jianhe Zheng and Ehud Nakar for stimulating discussions. O.G. is supported by the Flatiron Research Fellowship. The Flatiron Institute is supported by the Simons Foundation.

\bibliography{refs}

\begin{thebibliography}{}
\expandafter\ifx\csname natexlab\endcsname\relax\def\natexlab#1{#1}\fi
\providecommand{\url}[1]{\href{#1}{#1}}
\providecommand{\dodoi}[1]{doi:~\href{http://doi.org/#1}{\nolinkurl{#1}}}
\providecommand{\doeprint}[1]{\href{http://ascl.net/#1}{\nolinkurl{http://ascl.net/#1}}}
\providecommand{\doarXiv}[1]{\href{https://arxiv.org/abs/#1}{\nolinkurl{https://arxiv.org/abs/#1}}}

\bibitem[{{Alexander} {et~al.}(2020){Alexander}, {van Velzen}, {Horesh}, \& {Zauderer}}]{Alexander2020}
{Alexander}, K.~D., {van Velzen}, S., {Horesh}, A., \& {Zauderer}, B.~A. 2020, \ssr, 216, 81, \dodoi{10.1007/s11214-020-00702-w}

\bibitem[{{Aloy} \& {Obergaulinger}(2021)}]{Aloy&Obergaulinger2021}
{Aloy}, M.~{\'A}., \& {Obergaulinger}, M. 2021, \mnras, 500, 4365, \dodoi{10.1093/mnras/staa3273}

\bibitem[{{Aryan} {et~al.}(2025){Aryan}, {Chen}, {Yang}, {Gillanders}, {Kong}, {Smartt}, {Stevance}, {Yang}, {Aamer}, {Gupta}, {Fan}, {Hou}, {Hsiao}, {Kumar}, {Lai}, {Lee}, {Lee}, {Lin}, {Lin}, {Ngeow}, {Nicholl}, {Pan}, {Bhushan Pandey}, {Sankar. K}, {Srivastav}, {Sun}, \& {Wang}}]{Aryan2025}
{Aryan}, A., {Chen}, T.-W., {Yang}, S., {et~al.} 2025, arXiv e-prints, arXiv:2504.21096, \dodoi{10.48550/arXiv.2504.21096}

\bibitem[{{Ashall} {et~al.}(2019){Ashall}, {Mazzali}, {Pian}, {Woosley}, {Palazzi}, {Prentice}, {Kobayashi}, {Holmbo}, {Levan}, {Perley}, {Stritzinger}, {Bufano}, {Filippenko}, {Melandri}, {Oates}, {Rossi}, {Selsing}, {Zheng}, {Castro-Tirado}, {Chincarini}, {D'Avanzo}, {De Pasquale}, {Emery}, {Fruchter}, {Hurley}, {Moller}, {Nomoto}, {Tanaka}, \& {Valeev}}]{Ashall2019}
{Ashall}, C., {Mazzali}, P.~A., {Pian}, E., {et~al.} 2019, \mnras, 487, 5824, \dodoi{10.1093/mnras/stz1588}

\bibitem[{{Beniamini} {et~al.}(2020){Beniamini}, {Granot}, \& {Gill}}]{Beniamini2020}
{Beniamini}, P., {Granot}, J., \& {Gill}, R. 2020, \mnras, 493, 3521, \dodoi{10.1093/mnras/staa538}

\bibitem[{{Beniamini} {et~al.}(2016){Beniamini}, {Nava}, \& {Piran}}]{Beniamini2016}
{Beniamini}, P., {Nava}, L., \& {Piran}, T. 2016, \mnras, 461, 51, \dodoi{10.1093/mnras/stw1331}

\bibitem[{{Blandford} \& {Payne}(1982)}]{Blandford&Payne1982}
{Blandford}, R.~D., \& {Payne}, D.~G. 1982, \mnras, 199, 883, \dodoi{10.1093/mnras/199.4.883}

\bibitem[{{Blandford} \& {Znajek}(1977)}]{Blandford&Znajek1977}
{Blandford}, R.~D., \& {Znajek}, R.~L. 1977, \mnras, 179, 433, \dodoi{10.1093/mnras/179.3.433}

\bibitem[{{Bopp} \& {Gottlieb}(2025)}]{Bopp2025}
{Bopp}, J., \& {Gottlieb}, O. 2025, \apjl, 982, L56, \dodoi{10.3847/2041-8213/adbdcd}

\bibitem[{{Bright} {et~al.}(2022){Bright}, {Margutti}, {Matthews}, {Brethauer}, {Coppejans}, {Wieringa}, {Metzger}, {DeMarchi}, {Laskar}, {Romero}, {Alexander}, {Horesh}, {Migliori}, {Chornock}, {Berger}, {Bietenholz}, {Devlin}, {Dicker}, {Jacobson-Gal{\'a}n}, {Mason}, {Milisavljevic}, {Motta}, {Mroczkowski}, {Ramirez-Ruiz}, {Rhodes}, {Sarazin}, {Sfaradi}, \& {Sievers}}]{Bright2021}
{Bright}, J.~S., {Margutti}, R., {Matthews}, D., {et~al.} 2022, \apj, 926, 112, \dodoi{10.3847/1538-4357/ac4506}

\bibitem[{{Bright} {et~al.}(2025){Bright}, {Carotenuto}, {Fender}, {Choza}, {Mummery}, {Jonker}, {Smartt}, {DeBoer}, {Farah}, {Matthews}, {Pollak}, {Rhodes}, \& {Siemion}}]{Bright2025}
{Bright}, J.~S., {Carotenuto}, F., {Fender}, R., {et~al.} 2025, \apj, 981, 48, \dodoi{10.3847/1538-4357/adaaef}

\bibitem[{{Bromberg} {et~al.}(2011){Bromberg}, {Nakar}, {Piran}, \& {Sari}}]{Bromberg2011b}
{Bromberg}, O., {Nakar}, E., {Piran}, T., \& {Sari}, R. 2011, \apj, 740, 100, \dodoi{10.1088/0004-637X/740/2/100}

\bibitem[{{Bucciantini} {et~al.}(2009){Bucciantini}, {Quataert}, {Metzger}, {Thompson}, {Arons}, \& {Del Zanna}}]{Bucciantini2009}
{Bucciantini}, N., {Quataert}, E., {Metzger}, B.~D., {et~al.} 2009, \mnras, 396, 2038, \dodoi{10.1111/j.1365-2966.2009.14940.x}

\bibitem[{{Busmann} {et~al.}(2025){Busmann}, {O'Connor}, {Sommer}, {Gruen}, {Beniamini}, {Gill}, {Moss}, {Palmese}, {Riffeser}, {Yang}, {Troja}, {Dichiara}, {Ricci}, {Klingler}, {G{\"o}ssl}, {Hu}, {Rau}, {Ries}, {Ryan}, {Schmidt}, {Yadav}, \& {Zeimann}}]{Busmann2025}
{Busmann}, M., {O'Connor}, B., {Sommer}, J., {et~al.} 2025, arXiv e-prints, arXiv:2503.14588, \dodoi{10.48550/arXiv.2503.14588}

\bibitem[{{Cano}(2013)}]{Cano2013}
{Cano}, Z. 2013, \mnras, 434, 1098, \dodoi{10.1093/mnras/stt1048}

\bibitem[{{Chatterjee} \& {Narayan}(2022)}]{Chatterjee2022}
{Chatterjee}, K., \& {Narayan}, R. 2022, \apj, 941, 30, \dodoi{10.3847/1538-4357/ac9d97}

\bibitem[{{Chen} {et~al.}(2023){Chen}, {Drout}, {Piro}, {Kilpatrick}, {Foley}, {Rojas-Bravo}, {Taggart}, {Siebert}, \& {Magee}}]{Chen2023}
{Chen}, Y., {Drout}, M.~R., {Piro}, A.~L., {et~al.} 2023, \apj, 955, 42, \dodoi{10.3847/1538-4357/ace965}

\bibitem[{{Coppejans} {et~al.}(2020){Coppejans}, {Margutti}, {Terreran}, {Nayana}, {Coughlin}, {Laskar}, {Alexander}, {Bietenholz}, {Caprioli}, {Chandra}, {Drout}, {Frederiks}, {Frohmaier}, {Hurley}, {Kochanek}, {MacLeod}, {Meisner}, {Nugent}, {Ridnaia}, {Sand}, {Svinkin}, {Ward}, {Yang}, {Baldeschi}, {Chilingarian}, {Dong}, {Esquivia}, {Fong}, {Guidorzi}, {Lundqvist}, {Milisavljevic}, {Paterson}, {Reichart}, {Shappee}, {Stroh}, {Valenti}, {Zauderer}, \& {Zhang}}]{Coppejans2020}
{Coppejans}, D.~L., {Margutti}, R., {Terreran}, G., {et~al.} 2020, \apjl, 895, L23, \dodoi{10.3847/2041-8213/ab8cc7}

\bibitem[{{Dean} \& {Fern{\'a}ndez}(2024{\natexlab{a}})}]{Dean2024a}
{Dean}, C., \& {Fern{\'a}ndez}, R. 2024{\natexlab{a}}, \prd, 109, 083010, \dodoi{10.1103/PhysRevD.109.083010}

\bibitem[{{Dean} \& {Fern{\'a}ndez}(2024{\natexlab{b}})}]{Dean2024b}
---. 2024{\natexlab{b}}, \prd, 110, 083024, \dodoi{10.1103/PhysRevD.110.083024}

\bibitem[{{DeLaunay} {et~al.}(2024){DeLaunay}, {Tohuvavohu}, {Svinkin}, {Ronchini}, {Raman}, {Kennea}, \& {Parsotan}}]{DeLaunay2024}
{DeLaunay}, J., {Tohuvavohu}, A., {Svinkin}, D., {et~al.} 2024, GRB Coordinates Network, 35971, 1

\bibitem[{{Duffell} \& {Ho}(2020)}]{Duffell2020}
{Duffell}, P.~C., \& {Ho}, A. Y.~Q. 2020, \apj, 900, 193, \dodoi{10.3847/1538-4357/aba90a}

\bibitem[{{Duncan} \& {Thompson}(1992)}]{Duncan1992}
{Duncan}, R.~C., \& {Thompson}, C. 1992, \apjl, 392, L9, \dodoi{10.1086/186413}

\bibitem[{{Eisenberg} {et~al.}(2022){Eisenberg}, {Gottlieb}, \& {Nakar}}]{Eisenberg2022}
{Eisenberg}, M., {Gottlieb}, O., \& {Nakar}, E. 2022, \mnras, 517, 582, \dodoi{10.1093/mnras/stac2184}

\bibitem[{{Eyles-Ferris} {et~al.}(2025){Eyles-Ferris}, {Jonker}, {Levan}, {Malesani}, {Sarin}, {Fryer}, {Rastinejad}, {Burns}, {Tanvir}, {O'Brien}, {Fong}, {Mandel}, {Gompertz}, {Kilpatrick}, {Bloemen}, {Bright}, {Carotenuto}, {Corcoran}, {Cotter}, {Groot}, {Izzo}, {Laskar}, {Martin-Carrillo}, {Palmerio}, {Ravasio}, {van Roestel}, {Saccardi}, {Starling}, {Thakur}, {Vergani}, {Vreeswijk}, {Bauer}, {Campana}, {Chac{\'o}n}, {Chrimes}, {Covino}, {van Dalen}, {D'Elia}, {De Pasquale}, {Habeeb}, {Hartmann}, {van Hoof}, {Jakobsson}, {Julakanti}, {Leloudas}, {Mata S{\'a}nchez}, {Nixon}, {Pieterse}, {Pugliese}, {Quirola-V{\'a}squez}, {Rayson}, {Salvaterra}, {Schneider}, {Torres}, \& {Zafar}}]{EylesFerris2025}
{Eyles-Ferris}, R. A.~J., {Jonker}, P.~G., {Levan}, A.~J., {et~al.} 2025, \apjl, 988, L14, \dodoi{10.3847/2041-8213/ade1d9}

\bibitem[{{Fryer} {et~al.}(2025){Fryer}, {Burns}, {Ho}, {Corsi}, {Lien}, {Perley}, {Vail}, \& {Villar}}]{Fryer2025}
{Fryer}, C.~L., {Burns}, E., {Ho}, A. Y.~Q., {et~al.} 2025, \apj, 986, 185, \dodoi{10.3847/1538-4357/add474}

\bibitem[{{Fujibayashi} {et~al.}(2024){Fujibayashi}, {Lam}, {Shibata}, \& {Sekiguchi}}]{Fujibayashi2024}
{Fujibayashi}, S., {Lam}, A. T.-L., {Shibata}, M., \& {Sekiguchi}, Y. 2024, \prd, 109, 023031, \dodoi{10.1103/PhysRevD.109.023031}

\bibitem[{{Galama} {et~al.}(1998){Galama}, {Vreeswijk}, {van Paradijs}, {Kouveliotou}, {Augusteijn}, {B{\"o}hnhardt}, {Brewer}, {Doublier}, {Gonzalez}, {Leibundgut}, {Lidman}, {Hainaut}, {Patat}, {Heise}, {in't Zand}, {Hurley}, {Groot}, {Strom}, {Mazzali}, {Iwamoto}, {Nomoto}, {Umeda}, {Nakamura}, {Young}, {Suzuki}, {Shigeyama}, {Koshut}, {Kippen}, {Robinson}, {de Wildt}, {Wijers}, {Tanvir}, {Greiner}, {Pian}, {Palazzi}, {Frontera}, {Masetti}, {Nicastro}, {Feroci}, {Costa}, {Piro}, {Peterson}, {Tinney}, {Boyle}, {Cannon}, {Stathakis}, {Sadler}, {Begam}, \& {Ianna}}]{Galama1998}
{Galama}, T.~J., {Vreeswijk}, P.~M., {van Paradijs}, J., {et~al.} 1998, \nat, 395, 670, \dodoi{10.1038/27150}

\bibitem[{{Gianfagna} {et~al.}(2025){Gianfagna}, {Piro}, {Bruni}, {Linesh Thakur}, {Van Eerten}, {Castro-Tirado}, {Chen}, {Cheng}, {He}, {Jia}, {Ling}, {Maiorano}, {Paladino}, {Tripodi}, {Rossi}, {Yang}, {Yuan}, {Yuan}, \& {Zhang}}]{Gianfagna2025}
{Gianfagna}, G., {Piro}, L., {Bruni}, G., {et~al.} 2025, arXiv e-prints, arXiv:2505.05444, \dodoi{10.48550/arXiv.2505.05444}

\bibitem[{{Gillanders} {et~al.}(2024){Gillanders}, {Rhodes}, {Srivastav}, {Carotenuto}, {Bright}, {Huber}, {Stevance}, {Smartt}, {Chambers}, {Chen}, {Fender}, {Andersson}, {Cooper}, {Jonker}, {Cowie}, {de Boer}, {Erasmus}, {Fulton}, {Gao}, {Herman}, {Lin}, {Lowe}, {Magnier}, {Miao}, {Minguez}, {Moore}, {Ngeow}, {Nicholl}, {Pan}, {Pignata}, {Rest}, {Sheng}, {Smith}, {Smith}, {Tonry}, {Wainscoat}, {Weston}, {Yang}, \& {Young}}]{Gillanders2024}
{Gillanders}, J.~H., {Rhodes}, L., {Srivastav}, S., {et~al.} 2024, \apjl, 969, L14, \dodoi{10.3847/2041-8213/ad55cd}

\bibitem[{{Gottlieb} {et~al.}(2023{\natexlab{a}}){Gottlieb}, {Jacquemin-Ide}, {Lowell}, {Tchekhovskoy}, \& {Ramirez-Ruiz}}]{Gottlieb2023}
{Gottlieb}, O., {Jacquemin-Ide}, J., {Lowell}, B., {Tchekhovskoy}, A., \& {Ramirez-Ruiz}, E. 2023{\natexlab{a}}, \apjl, 952, L32, \dodoi{10.3847/2041-8213/ace779}

\bibitem[{{Gottlieb} {et~al.}(2022{\natexlab{a}}){Gottlieb}, {Lalakos}, {Bromberg}, {Liska}, \& {Tchekhovskoy}}]{Gottlieb2022a}
{Gottlieb}, O., {Lalakos}, A., {Bromberg}, O., {Liska}, M., \& {Tchekhovskoy}, A. 2022{\natexlab{a}}, \mnras, 510, 4962, \dodoi{10.1093/mnras/stab3784}

\bibitem[{{Gottlieb} {et~al.}(2019{\natexlab{a}}){Gottlieb}, {Levinson}, \& {Nakar}}]{Gottlieb2019b}
{Gottlieb}, O., {Levinson}, A., \& {Nakar}, E. 2019{\natexlab{a}}, \mnras, 488, 1416, \dodoi{10.1093/mnras/stz1828}

\bibitem[{{Gottlieb} {et~al.}(2020){Gottlieb}, {Levinson}, \& {Nakar}}]{Gottlieb2020c}
---. 2020, \mnras, 495, 570, \dodoi{10.1093/mnras/staa1216}

\bibitem[{{Gottlieb} {et~al.}(2022{\natexlab{b}}){Gottlieb}, {Liska}, {Tchekhovskoy}, {Bromberg}, {Lalakos}, {Giannios}, \& {M{\"o}sta}}]{Gottlieb2022c}
{Gottlieb}, O., {Liska}, M., {Tchekhovskoy}, A., {et~al.} 2022{\natexlab{b}}, \apjl, 933, L9, \dodoi{10.3847/2041-8213/ac7530}

\bibitem[{{Gottlieb} {et~al.}(2025){Gottlieb}, {Metzger}, {Issa}, {Li}, {Renzo}, \& {Isi}}]{Gottlieb2025}
{Gottlieb}, O., {Metzger}, B.~D., {Issa}, D., {et~al.} 2025, arXiv e-prints, arXiv:2508.15887, \dodoi{10.48550/arXiv.2508.15887}

\bibitem[{{Gottlieb} {et~al.}(2022{\natexlab{c}}){Gottlieb}, {Moseley}, {Ramirez-Aguilar}, {Murguia-Berthier}, {Liska}, \& {Tchekhovskoy}}]{Gottlieb2022d}
{Gottlieb}, O., {Moseley}, S., {Ramirez-Aguilar}, T., {et~al.} 2022{\natexlab{c}}, \apjl, 933, L2, \dodoi{10.3847/2041-8213/ac7728}

\bibitem[{{Gottlieb} {et~al.}(2021){Gottlieb}, {Nakar}, \& {Bromberg}}]{Gottlieb2021a}
{Gottlieb}, O., {Nakar}, E., \& {Bromberg}, O. 2021, \mnras, 500, 3511, \dodoi{10.1093/mnras/staa3501}

\bibitem[{{Gottlieb} {et~al.}(2019{\natexlab{b}}){Gottlieb}, {Nakar}, \& {Piran}}]{Gottlieb2019a}
{Gottlieb}, O., {Nakar}, E., \& {Piran}, T. 2019{\natexlab{b}}, \mnras, 488, 2405, \dodoi{10.1093/mnras/stz1906}

\bibitem[{{Gottlieb} {et~al.}(2024){Gottlieb}, {Renzo}, {Metzger}, {Goldberg}, \& {Cantiello}}]{Gottlieb2024b}
{Gottlieb}, O., {Renzo}, M., {Metzger}, B.~D., {Goldberg}, J.~A., \& {Cantiello}, M. 2024, \apjl, 976, L13, \dodoi{10.3847/2041-8213/ad8563}

\bibitem[{{Gottlieb} {et~al.}(2022{\natexlab{d}}){Gottlieb}, {Tchekhovskoy}, \& {Margutti}}]{Gottlieb2022b}
{Gottlieb}, O., {Tchekhovskoy}, A., \& {Margutti}, R. 2022{\natexlab{d}}, \mnras, 513, 3810, \dodoi{10.1093/mnras/stac910}

\bibitem[{{Gottlieb} {et~al.}(2023{\natexlab{b}}){Gottlieb}, {Issa}, {Jacquemin-Ide}, {Liska}, {Foucart}, {Tchekhovskoy}, {Metzger}, {Quataert}, {Perna}, {Kasen}, {Duez}, {Kidder}, {Pfeiffer}, \& {Scheel}}]{Gottlieb2023c}
{Gottlieb}, O., {Issa}, D., {Jacquemin-Ide}, J., {et~al.} 2023{\natexlab{b}}, \apjl, 954, L21, \dodoi{10.3847/2041-8213/aceeff}

\bibitem[{{Granot}(2005)}]{Granot2005}
{Granot}, J. 2005, \apj, 631, 1022, \dodoi{10.1086/432676}

\bibitem[{{Granot} {et~al.}(2003){Granot}, {Nakar}, \& {Piran}}]{Granot2003}
{Granot}, J., {Nakar}, E., \& {Piran}, T. 2003, \nat, 426, 138, \dodoi{10.1038/426138a}

\bibitem[{{Granot} {et~al.}(2002){Granot}, {Panaitescu}, {Kumar}, \& {Woosley}}]{Granot2002}
{Granot}, J., {Panaitescu}, A., {Kumar}, P., \& {Woosley}, S.~E. 2002, \apjl, 570, L61, \dodoi{10.1086/340991}

\bibitem[{{Guti{\'e}rrez} {et~al.}(2024){Guti{\'e}rrez}, {Mattila}, {Lundqvist}, {Dessart}, {Gonz{\'a}lez-Gait{\'a}n}, {Jonker}, {Dong}, {Coppejans}, {Chen}, {Charalampopoulos}, {Elias-Rosa}, {Reynolds}, {Kochanek}, {Fraser}, {Pastorello}, {Gromadzki}, {Neustadt}, {Benetti}, {Kankare}, {Kangas}, {Kotak}, {Stritzinger}, {Wevers}, {Zhang}, {Bersier}, {Bose}, {Buckley}, {Dastidar}, {Gangopadhyay}, {Hamanowicz}, {Kollmeier}, {Mao}, {Misra}, {Potter}, {Prieto}, {Romero-Colmenero}, {Singh}, {Somero}, {Terreran}, {Vaisanen}, \& {Wyrzykowski}}]{Gutierrez2024}
{Guti{\'e}rrez}, C.~P., {Mattila}, S., {Lundqvist}, P., {et~al.} 2024, \apj, 977, 162, \dodoi{10.3847/1538-4357/ad89a5}

\bibitem[{{Hamidani} {et~al.}(2025{\natexlab{a}}){Hamidani}, {Ioka}, {Kashiyama}, \& {Tanaka}}]{Hamidani2025a}
{Hamidani}, H., {Ioka}, K., {Kashiyama}, K., \& {Tanaka}, M. 2025{\natexlab{a}}, \apj, 988, 30, \dodoi{10.3847/1538-4357/addd13}

\bibitem[{{Hamidani} {et~al.}(2025{\natexlab{b}}){Hamidani}, {Sato}, {Kashiyama}, {Tanaka}, {Ioka}, \& {Kimura}}]{Hamidani2025}
{Hamidani}, H., {Sato}, Y., {Kashiyama}, K., {et~al.} 2025{\natexlab{b}}, \apjl, 986, L4, \dodoi{10.3847/2041-8213/add99d}

\bibitem[{{Harrison} {et~al.}(2018){Harrison}, {Gottlieb}, \& {Nakar}}]{Harrison2018}
{Harrison}, R., {Gottlieb}, O., \& {Nakar}, E. 2018, \mnras, 477, 2128, \dodoi{10.1093/mnras/sty760}

\bibitem[{{Hayakawa} \& {Maeda}(2018)}]{Hayakawa2018}
{Hayakawa}, T., \& {Maeda}, K. 2018, \apj, 854, 43, \dodoi{10.3847/1538-4357/aaa76c}

\bibitem[{{Ho} {et~al.}(2019){Ho}, {Phinney}, {Ravi}, {Kulkarni}, {Petitpas}, {Emonts}, {Bhalerao}, {Blundell}, {Cenko}, {Dobie}, {Howie}, {Kamraj}, {Kasliwal}, {Murphy}, {Perley}, {Sridharan}, \& {Yoon}}]{Ho2019}
{Ho}, A. Y.~Q., {Phinney}, E.~S., {Ravi}, V., {et~al.} 2019, \apj, 871, 73, \dodoi{10.3847/1538-4357/aaf473}

\bibitem[{{Ho} {et~al.}(2020){Ho}, {Kulkarni}, {Perley}, {Cenko}, {Corsi}, {Schulze}, {Lunnan}, {Sollerman}, {Gal-Yam}, {Anand}, {Barbarino}, {Bellm}, {Bruch}, {Burns}, {De}, {Dekany}, {Delacroix}, {Duev}, {Frederiks}, {Fremling}, {Goldstein}, {Golkhou}, {Graham}, {Hale}, {Kasliwal}, {Kupfer}, {Laher}, {Martikainen}, {Masci}, {Neill}, {Ridnaia}, {Rusholme}, {Savchenko}, {Shupe}, {Soumagnac}, {Strotjohann}, {Svinkin}, {Taggart}, {Tartaglia}, {Yan}, \& {Zolkower}}]{Ho2020}
{Ho}, A. Y.~Q., {Kulkarni}, S.~R., {Perley}, D.~A., {et~al.} 2020, \apj, 902, 86, \dodoi{10.3847/1538-4357/aba630}

\bibitem[{{Ho} {et~al.}(2022){Ho}, {Margalit}, {Bremer}, {Perley}, {Yao}, {Dobie}, {Kaplan}, {O'Brien}, {Petitpas}, \& {Zic}}]{Ho2022}
{Ho}, A. Y.~Q., {Margalit}, B., {Bremer}, M., {et~al.} 2022, \apj, 932, 116, \dodoi{10.3847/1538-4357/ac4e97}

\bibitem[{{Issa} {et~al.}(2025){Issa}, {Gottlieb}, {Metzger}, {Jacquemin-Ide}, {Liska}, {Foucart}, {Halevi}, \& {Tchekhovskoy}}]{Issa2025}
{Issa}, D., {Gottlieb}, O., {Metzger}, B.~D., {et~al.} 2025, \apjl, 985, L26, \dodoi{10.3847/2041-8213/adc694}

\bibitem[{{Ito} {et~al.}(2015){Ito}, {Matsumoto}, {Nagataki}, {Warren}, \& {Barkov}}]{Ito2015}
{Ito}, H., {Matsumoto}, J., {Nagataki}, S., {Warren}, D.~C., \& {Barkov}, M.~V. 2015, \apjl, 814, L29, \dodoi{10.1088/2041-8205/814/2/L29}

\bibitem[{{Izzo} {et~al.}(2025){Izzo}, {Martin-Carrillo}, {Malesani}, {Levan}, {Jonker}, {Cotter}, {van Dalen}, {Corcoran}, {Wiersema}, \& {Bauer}}]{Izzo2025}
{Izzo}, L., {Martin-Carrillo}, A., {Malesani}, D.~B., {et~al.} 2025, GRB Coordinates Network, 39851, 1

\bibitem[{{Jiang} {et~al.}(2025){Jiang}, {Xu}, {van Hoof}, {Lei}, {Liu}, {Zhou}, {Chen}, {Fu}, {Yang}, {Liu}, {Zhu}, {Filippenko}, {Jonker}, {Pozanenko}, {Gao}, {Wu}, {Zhang}, {Lamb}, {De Pasquale}, {Kobayashi}, {Bauer}, {Sun}, {Pugliese}, {An}, {D'Elia}, {Fynbo}, {Zheng}, {Tirado}, {Yin}, {Zou}, {Deller}, {Pankov}, {Volnova}, {Moskvitin}, {Spiridonova}, {Oparin}, {Rumyantsev}, {Burkhonov}, {Egamberdiyev}, {Kim}, {Krugov}, {Tatarnikov}, {Inasaridze}, {Levan}, {Bj{\o}rn Malesani}, {Ravasio}, {Quirola-V{\'a}squez}, {van Dalen}, {S{\'a}nchez-Sierras}, {Mata S{\'a}nchez}, {Littlefair}, {Chac{\'o}n}, {Torres}, {Chrimes}, {Sarin}, {Martin-Carrillo}, {Dhillon}, {Yang}, {Brink}, {Davies}, {Yang}, {Aryan}, {Chen}, {Kong}, {Li}, {Li}, {Mao}, {P{\'e}rez-Garc{\'\i}a}, {Fern{\'a}ndez-Garc{\'\i}a}, {Andrews}, {Farah}, {Fan}, {Padilla Gonzalez}, {Howell}, {Hartmann}, {Hu}, {Jakobsson}, {Li}, {Ling}, {McCully}, {Newsome}, {Schneider}, {Samaporn Tinyanont}, {Sun}, {Terreran}, {Tang}, {Wang}, {Xu}, {Yuan}, {Zhang}, {Zhao}, \&
  {Zhang}}]{Jiang2025}
{Jiang}, S.-Q., {Xu}, D., {van Hoof}, A. P.~C., {et~al.} 2025, arXiv e-prints, arXiv:2503.04306, \dodoi{10.48550/arXiv.2503.04306}

\bibitem[{{Kouveliotou} {et~al.}(1993){Kouveliotou}, {Meegan}, {Fishman}, {Bhat}, {Briggs}, {Koshut}, {Paciesas}, \& {Pendleton}}]{Kouveliotou1993}
{Kouveliotou}, C., {Meegan}, C.~A., {Fishman}, G.~J., {et~al.} 1993, \apjl, 413, L101, \dodoi{10.1086/186969}

\bibitem[{{Kuroda} {et~al.}(2020){Kuroda}, {Arcones}, {Takiwaki}, \& {Kotake}}]{Kuroda2020}
{Kuroda}, T., {Arcones}, A., {Takiwaki}, T., \& {Kotake}, K. 2020, \apj, 896, 102, \dodoi{10.3847/1538-4357/ab9308}

\bibitem[{{Lamb} {et~al.}(2021){Lamb}, {Fern{\'a}ndez}, {Hayes}, {Kong}, {Lin}, {Tanvir}, {Hendry}, {Heng}, {Saha}, \& {Veitch}}]{Lamb2021}
{Lamb}, G.~P., {Fern{\'a}ndez}, J.~J., {Hayes}, F., {et~al.} 2021, Universe, 7, 329, \dodoi{10.3390/universe7090329}

\bibitem[{{Lazzati} \& {Begelman}(2005)}]{Lazzati2005}
{Lazzati}, D., \& {Begelman}, M.~C. 2005, \apj, 629, 903, \dodoi{10.1086/430877}

\bibitem[{{Levan} {et~al.}(2024){Levan}, {Jonker}, {Saccardi}, {Bj{\o}rn Malesani}, {Tanvir}, {Izzo}, {Heintz}, {Mata S{\'a}nchez}, {Quirola-V{\'a}squez}, {Torres}, {Vergani}, {Schulze}, {Rossi}, {D'Avanzo}, {Gompertz}, {Martin-Carrillo}, {de Ugarte Postigo}, {Schneider}, {Yuan}, {Ling}, {Zhang}, {Mao}, {Liu}, {Sun}, {Xu}, {Zhu}, {Ag{\"u}{\'\i} Fern{\'a}ndez}, {Amati}, {Bauer}, {Campana}, {Carotenuto}, {Chrimes}, {van Dalen}, {D'Elia}, {Della Valle}, {De Pasquale}, {Dhillon}, {Galbany}, {Gaspari}, {Gianfagna}, {Gomboc}, {Habeeb}, {van Hoof}, {Hu}, {Jakobsson}, {Julakanti}, {Korth}, {Kouveliotou}, {Laskar}, {Littlefair}, {Maiorano}, {Mao}, {Melandri}, {Miller}, {Mukherjee}, {Oates}, {O'Brien}, {Palmerio}, {Parviainen}, {Pieterse}, {Piranomonte}, {Piro}, {Pugliese}, {Ravasio}, {Rayson}, {Salvaterra}, {S{\'a}nchez-Ram{\'\i}rez}, {Sarin}, {Shilling}, {Starling}, {Tagliaferri}, {Linesh Thakur}, {Th{\"o}ne}, {Wiersema}, {Worssam}, \& {Zafar}}]{Levan2024}
{Levan}, A.~J., {Jonker}, P.~G., {Saccardi}, A., {et~al.} 2024, arXiv e-prints, arXiv:2404.16350, \dodoi{10.48550/arXiv.2404.16350}

\bibitem[{{Li} {et~al.}(2023){Li}, {Gao}, {Ai}, \& {Lei}}]{Li2023}
{Li}, J.-D., {Gao}, H., {Ai}, S., \& {Lei}, W.-H. 2023, \mnras, 525, 6285, \dodoi{10.1093/mnras/stad2606}

\bibitem[{{Li} {et~al.}(2025){Li}, {Zhu}, {Zou}, {Geng}, {Liu}, {Wang}, {Li}, {Xu}, {Sun}, {Wang}, {Yu}, {Zhang}, {Wu}, {Yang}, {Filippenko}, {Liu}, {Yuan}, {Aguado}, {An}, {An}, {Buckley}, {Castro-Tirado}, {Fu}, {Fynbo}, {Howell}, {Hu}, {Jiang}, {Kumar}, {Mao}, {Maund}, {Liu}, {Mockler}, {Moskvitin}, {Andrews}, {Bom}, {Brink}, {Chatterjee}, {Chen}, {Cheng}, {Cooke}, {Dai}, {Du}, {Erasmus}, {Fang}, {Farah}, {Goranskij}, {Gritsevich}, {Gu}, {Guo}, {Hsiao}, {Hu}, {Hua}, {Jacobson-Gal{\'a}n}, {Jia}, {Jin}, {Kasliwal}, {Kilpatrick}, {Kumar}, {Lei}, {Li}, {Li}, {Li}, {Ling}, {Liu}, {Liu}, {Liu}, {L{\'o}pez-Oramas}, {Maslennikova}, {McCully}, {Monageng}, {Newsone}, {Padilla Gonzalez}, {Pan}, {Peng}, {Pignata}, {Poidevin}, {Potter}, {P{\'e}rez-Fournon}, {Santana-Silva}, {Santos}, {Song}, {Song}, {Spiridonova}, {Sun}, {Sun}, {Terreran}, {Wang}, {Wang}, {Wang}, {Wang}, {Wu}, {Xiang}, {Xiao}, {Xu}, {Xue}, {Yan}, {Yang}, {Yu}, {Zhang}, {Zhang}, {Zhang}, {Zhang}, {Zhang}, {Zheng}, \& {Zou}}]{Li2025}
{Li}, W.~X., {Zhu}, Z.~P., {Zou}, X.~Z., {et~al.} 2025, arXiv e-prints, arXiv:2504.17034, \dodoi{10.48550/arXiv.2504.17034}

\bibitem[{{Liang} {et~al.}(2013){Liang}, {Li}, {Gao}, {Zhang}, {Liang}, {Wu}, {Yi}, {Dai}, {Tang}, {Chen}, {L{\"u}}, {Zhang}, {Lu}, {L{\"u}}, \& {Wei}}]{Liang2013}
{Liang}, E.-W., {Li}, L., {Gao}, H., {et~al.} 2013, \apj, 774, 13, \dodoi{10.1088/0004-637X/774/1/13}

\bibitem[{{Liu} {et~al.}(2025){Liu}, {Sun}, {Xu}, {Svinkin}, {Delaunay}, {Tanvir}, {Gao}, {Zhang}, {Chen}, {Wu}, {Zhang}, {Yuan}, {An}, {Bruni}, {Frederiks}, {Ghirlanda}, {Hu}, {Li}, {Li}, {Li}, {Malesani}, {Piro}, {Raman}, {Ricci}, {Troja}, {Vergani}, {Wu}, {Yang}, {Zhang}, {Zhu}, {de Ugarte Postigo}, {Demin}, {Dobie}, {Fan}, {Fu}, {Fynbo}, {Geng}, {Gianfagna}, {Hu}, {Huang}, {Jiang}, {Jonker}, {Julakanti}, {Kennea}, {Kokomov}, {Kuulkers}, {Lei}, {Leung}, {Levan}, {Li}, {Li}, {Littlefair}, {Liu}, {Lysenko}, {Ma}, {Martin-Carrillo}, {O'Brien}, {Parsotan}, {Quirola-V{\'a}squez}, {Ridnaia}, {Ronchini}, {Rossi}, {Mata-S{\'a}nchez}, {Schneider}, {Shen}, {Thakur}, {Tohuvavohu}, {Torres}, {Tsvetkova}, {Ulanov}, {Wei}, {Xiao}, {Yin}, {Bai}, {Burwitz}, {Cai}, {Chen}, {Chen}, {Chen}, {Chen}, {Chen}, {Chen}, {Cheng}, {Cordier}, {Cui}, {Cui}, {Dai}, {Dai}, {Eder}, {Eyles-Ferris}, {Fan}, {Feldman}, {Feng}, {Feng}, {Friedrich}, {Gao}, {Gonzalez}, {Guan}, {Han}, {Han}, {Hou}, {Hu}, {Hu}, {Huang}, {Huo}, {Hutchinson}, {Ji},
  {Jia}, {Jia}, {Jiang}, {Jin}, {Jin}, {Jin}, {Keereman}, {Lerman}, {Li}, {Li}, {Li}, {Li}, {Li}, {Lian}, {Liang}, {Ling}, {Liu}, {Liu}, {Liu}, {Liu}, {Liu}, {Lu}, {L{\"u}}, {Luo}, {Ma}, {Ma}, {Mao}, {Mao}, {McHugh}, {Meidinger}, {Nandra}, {Osborne}, {Pan}, {Pan}, {Ravasio}, {Rau}, {Rea}, {Rehman}, {Sanders}, {Santovincenzo}, {Song}, {Su}, {Sun}, {Sun}, {Sun}, {Tan}, {Tang}, {Tao}, {Tong}, {Wang}, {Wang}, {Wang}, {Wang}, {Wang}, {Wang}, {Wang}, {Wang}, {Wang}, {Wei}, {Willingale}, {Xiong}, {Xu}, {Xu}, {Xu}, {Xu}, {Xu}, {Xue}, {Xue}, {Yan}, {Yang}, {Yang}, {Yang}, {Yang}, {Yu}, {Zhang}, {Zhang}, {Zhang}, {Zhang}, {Zhang}, {Zhang}, {Zhang}, {Zhang}, {Zhang}, {Zhao}, {Zhao}, {Zhao}, {Zhao}, {Zhou}, {Zhou}, {Zhu}, {Zhu}, \& {Zuo}}]{Liu2025}
{Liu}, Y., {Sun}, H., {Xu}, D., {et~al.} 2025, Nature Astronomy, 9, 564, \dodoi{10.1038/s41550-024-02449-8}

\bibitem[{{MacFadyen} \& {Woosley}(1999)}]{MacFadyen&Woosley1999}
{MacFadyen}, A.~I., \& {Woosley}, S.~E. 1999, \apj, 524, 262, \dodoi{10.1086/307790}

\bibitem[{{Maeda} {et~al.}(2003){Maeda}, {Mazzali}, {Deng}, {Nomoto}, {Yoshii}, {Tomita}, \& {Kobayashi}}]{Maeda2003b}
{Maeda}, K., {Mazzali}, P.~A., {Deng}, J., {et~al.} 2003, \apj, 593, 931, \dodoi{10.1086/376591}

\bibitem[{{Maeda} \& {Nomoto}(2003)}]{Maeda2003}
{Maeda}, K., \& {Nomoto}, K. 2003, \apj, 598, 1163, \dodoi{10.1086/378948}

\bibitem[{{Margutti} {et~al.}(2019){Margutti}, {Metzger}, {Chornock}, {Vurm}, {Roth}, {Grefenstette}, {Savchenko}, {Cartier}, {Steiner}, {Terreran}, {Margalit}, {Migliori}, {Milisavljevic}, {Alexander}, {Bietenholz}, {Blanchard}, {Bozzo}, {Brethauer}, {Chilingarian}, {Coppejans}, {Ducci}, {Ferrigno}, {Fong}, {G{\"o}tz}, {Guidorzi}, {Hajela}, {Hurley}, {Kuulkers}, {Laurent}, {Mereghetti}, {Nicholl}, {Patnaude}, {Ubertini}, {Banovetz}, {Bartel}, {Berger}, {Coughlin}, {Eftekhari}, {Frederiks}, {Kozlova}, {Laskar}, {Svinkin}, {Drout}, {MacFadyen}, \& {Paterson}}]{Margutti2019}
{Margutti}, R., {Metzger}, B.~D., {Chornock}, R., {et~al.} 2019, \apj, 872, 18, \dodoi{10.3847/1538-4357/aafa01}

\bibitem[{{Masada} {et~al.}(2022){Masada}, {Takiwaki}, \& {Kotake}}]{Masada2022}
{Masada}, Y., {Takiwaki}, T., \& {Kotake}, K. 2022, \apj, 924, 75, \dodoi{10.3847/1538-4357/ac34f6}

\bibitem[{{Matthews} {et~al.}(2023){Matthews}, {Margutti}, {Metzger}, {Milisavljevic}, {Migliori}, {Laskar}, {Brethauer}, {Berger}, {Chornock}, {Drout}, \& {Ramirez-Ruiz}}]{Matthews2023}
{Matthews}, D., {Margutti}, R., {Metzger}, B.~D., {et~al.} 2023, Research Notes of the American Astronomical Society, 7, 126, \dodoi{10.3847/2515-5172/acdde1}

\bibitem[{{Matzner}(2003)}]{Matzner2003}
{Matzner}, C.~D. 2003, \mnras, 345, 575, \dodoi{10.1046/j.1365-8711.2003.06969.x}

\bibitem[{McBreen {et~al.}(1994)McBreen, Hurley, Long, \& Metcalfe}]{McBreen1994}
McBreen, B., Hurley, K.~J., Long, R., \& Metcalfe, L. 1994, Monthly Notices of the Royal Astronomical Society, 271, 662, \dodoi{10.1093/mnras/271.3.662}

\bibitem[{{McKinney} {et~al.}(2012){McKinney}, {Tchekhovskoy}, \& {Blandford}}]{McKinney2012}
{McKinney}, J.~C., {Tchekhovskoy}, A., \& {Blandford}, R.~D. 2012, \mnras, 423, 3083, \dodoi{10.1111/j.1365-2966.2012.21074.x}

\bibitem[{{Menegazzi} {et~al.}(2025){Menegazzi}, {Fujibayashi}, {Shibata}, {Betranhandy}, \& {Takahashi}}]{Menegazzi2025}
{Menegazzi}, L.~C., {Fujibayashi}, S., {Shibata}, M., {Betranhandy}, A., \& {Takahashi}, K. 2025, \mnras, 537, 2850, \dodoi{10.1093/mnras/staf179}

\bibitem[{{Menegazzi} {et~al.}(2024){Menegazzi}, {Fujibayashi}, {Takahashi}, \& {Ishii}}]{Menegazzi2024}
{Menegazzi}, L.~C., {Fujibayashi}, S., {Takahashi}, K., \& {Ishii}, A. 2024, \mnras, 529, 178, \dodoi{10.1093/mnras/stae544}

\bibitem[{{M{\'e}sz{\'a}ros} \& {Rees}(2001)}]{Meszaros2001}
{M{\'e}sz{\'a}ros}, P., \& {Rees}, M.~J. 2001, \apjl, 556, L37, \dodoi{10.1086/322934}

\bibitem[{{Metzger} {et~al.}(2011){Metzger}, {Giannios}, {Thompson}, {Bucciantini}, \& {Quataert}}]{Metzger2011}
{Metzger}, B.~D., {Giannios}, D., {Thompson}, T.~A., {Bucciantini}, N., \& {Quataert}, E. 2011, \mnras, 413, 2031, \dodoi{10.1111/j.1365-2966.2011.18280.x}

\bibitem[{{Metzger} {et~al.}(2015){Metzger}, {Margalit}, {Kasen}, \& {Quataert}}]{Metzger2015}
{Metzger}, B.~D., {Margalit}, B., {Kasen}, D., \& {Quataert}, E. 2015, \mnras, 454, 3311, \dodoi{10.1093/mnras/stv2224}

\bibitem[{{Metzger} {et~al.}(2008){Metzger}, {Thompson}, \& {Quataert}}]{Metzger2008}
{Metzger}, B.~D., {Thompson}, T.~A., \& {Quataert}, E. 2008, \apj, 676, 1130, \dodoi{10.1086/526418}

\bibitem[{{Migliori} {et~al.}(2024){Migliori}, {Margutti}, {Metzger}, {Chornock}, {Vignali}, {Brethauer}, {Coppejans}, {Maccarone}, {Rivera Sandoval}, {Bright}, {Laskar}, {Milisavljevic}, {Berger}, \& {Nayana}}]{Migliori2024}
{Migliori}, G., {Margutti}, R., {Metzger}, B.~D., {et~al.} 2024, \apjl, 963, L24, \dodoi{10.3847/2041-8213/ad2764}

\bibitem[{{Mignone} {et~al.}(2007){Mignone}, {Bodo}, {Massaglia}, {Matsakos}, {Tesileanu}, {Zanni}, \& {Ferrari}}]{Mignone2007}
{Mignone}, A., {Bodo}, G., {Massaglia}, S., {et~al.} 2007, \apjs, 170, 228, \dodoi{10.1086/513316}

\bibitem[{{Minaev} \& {Pozanenko}(2020)}]{Minaev2020}
{Minaev}, P.~Y., \& {Pozanenko}, A.~S. 2020, \mnras, 492, 1919, \dodoi{10.1093/mnras/stz3611}

\bibitem[{{Mooley} {et~al.}(2018){Mooley}, {Deller}, {Gottlieb}, {Nakar}, {Hallinan}, {Bourke}, {Frail}, {Horesh}, {Corsi}, \& {Hotokezaka}}]{Mooley2018}
{Mooley}, K.~P., {Deller}, A.~T., {Gottlieb}, O., {et~al.} 2018, \nat, 561, 355, \dodoi{10.1038/s41586-018-0486-3}

\bibitem[{{Moriya} {et~al.}(2017){Moriya}, {Chen}, \& {Langer}}]{Moriya2017}
{Moriya}, T.~J., {Chen}, T.-W., \& {Langer}, N. 2017, \apj, 835, 177, \dodoi{10.3847/1538-4357/835/2/177}

\bibitem[{{Moriya} {et~al.}(2020){Moriya}, {Suzuki}, {Takiwaki}, {Pan}, \& {Blinnikov}}]{Moriya2020}
{Moriya}, T.~J., {Suzuki}, A., {Takiwaki}, T., {Pan}, Y.-C., \& {Blinnikov}, S.~I. 2020, \mnras, 497, 1619, \dodoi{10.1093/mnras/staa2060}

\bibitem[{{M{\"o}sta} {et~al.}(2015){M{\"o}sta}, {Ott}, {Radice}, {Roberts}, {Schnetter}, \& {Haas}}]{Mosta2015}
{M{\"o}sta}, P., {Ott}, C.~D., {Radice}, D., {et~al.} 2015, \nat, 528, 376, \dodoi{10.1038/nature15755}

\bibitem[{{M{\"o}sta} {et~al.}(2018){M{\"o}sta}, {Roberts}, {Halevi}, {Ott}, {Lippuner}, {Haas}, \& {Schnetter}}]{Mosta2018}
{M{\"o}sta}, P., {Roberts}, L.~F., {Halevi}, G., {et~al.} 2018, \apj, 864, 171, \dodoi{10.3847/1538-4357/aad6ec}

\bibitem[{{M{\"o}sta} {et~al.}(2014){M{\"o}sta}, {Richers}, {Ott}, {Haas}, {Piro}, {Boydstun}, {Abdikamalov}, {Reisswig}, \& {Schnetter}}]{Mosta2014}
{M{\"o}sta}, P., {Richers}, S., {Ott}, C.~D., {et~al.} 2014, \apjl, 785, L29, \dodoi{10.1088/2041-8205/785/2/L29}

\bibitem[{{Nakar}(2015)}]{Nakar2015}
{Nakar}, E. 2015, \apj, 807, 172, \dodoi{10.1088/0004-637X/807/2/172}

\bibitem[{{Nakar} {et~al.}(2018){Nakar}, {Gottlieb}, {Piran}, {Kasliwal}, \& {Hallinan}}]{Nakar2018}
{Nakar}, E., {Gottlieb}, O., {Piran}, T., {Kasliwal}, M.~M., \& {Hallinan}, G. 2018, \apj, 867, 18, \dodoi{10.3847/1538-4357/aae205}

\bibitem[{{Nakar} \& {Granot}(2007)}]{Nakar2007}
{Nakar}, E., \& {Granot}, J. 2007, \mnras, 380, 1744, \dodoi{10.1111/j.1365-2966.2007.12245.x}

\bibitem[{{Nakar} \& {Piran}(2017)}]{Nakar2017}
{Nakar}, E., \& {Piran}, T. 2017, \apj, 834, 28, \dodoi{10.3847/1538-4357/834/1/28}

\bibitem[{{Nakar} {et~al.}(2002){Nakar}, {Piran}, \& {Granot}}]{Nakar2002}
{Nakar}, E., {Piran}, T., \& {Granot}, J. 2002, \apj, 579, 699, \dodoi{10.1086/342791}

\bibitem[{{Narayan} {et~al.}(2012){Narayan}, {S{\"A} dowski}, {Penna}, \& {Kulkarni}}]{Narayan2012}
{Narayan}, R., {S{\"A} dowski}, A., {Penna}, R.~F., \& {Kulkarni}, A.~K. 2012, \mnras, 426, 3241, \dodoi{10.1111/j.1365-2966.2012.22002.x}

\bibitem[{{Nicholl} {et~al.}(2023){Nicholl}, {Srivastav}, {Fulton}, {Gomez}, {Huber}, {Oates}, {Ramsden}, {Rhodes}, {Smartt}, {Smith}, {Aamer}, {Anderson}, {Bauer}, {Berger}, {de Boer}, {Chambers}, {Charalampopoulos}, {Chen}, {Fender}, {Fraser}, {Gao}, {Green}, {Galbany}, {Gompertz}, {Gromadzki}, {Guti{\'e}rrez}, {Howell}, {Inserra}, {Jonker}, {Kopsacheili}, {Lowe}, {Magnier}, {McCully}, {McGee}, {Moore}, {M{\"u}ller-Bravo}, {Newsome}, {Gonzalez}, {Pellegrino}, {Pessi}, {Pursiainen}, {Rest}, {Ridley}, {Shappee}, {Sheng}, {Smith}, {Terreran}, {Tucker}, {Vink{\'o}}, {Wainscoat}, {Wiseman}, \& {Young}}]{Nicholl2023}
{Nicholl}, M., {Srivastav}, S., {Fulton}, M.~D., {et~al.} 2023, \apjl, 954, L28, \dodoi{10.3847/2041-8213/acf0ba}

\bibitem[{{Obergaulinger} \& {Aloy}(2022)}]{Obergaulinger&Aloy2022}
{Obergaulinger}, M., \& {Aloy}, M.~{\'A}. 2022, \mnras, 512, 2489, \dodoi{10.1093/mnras/stac613}

\bibitem[{{Panaitescu} {et~al.}(1998){Panaitescu}, {M{\'e}sz{\'a}ros}, \& {Rees}}]{Panaitescu1998}
{Panaitescu}, A., {M{\'e}sz{\'a}ros}, P., \& {Rees}, M.~J. 1998, \apj, 503, 314, \dodoi{10.1086/305995}

\bibitem[{{Perley} {et~al.}(2019){Perley}, {Mazzali}, {Yan}, {Cenko}, {Gezari}, {Taggart}, {Blagorodnova}, {Fremling}, {Mockler}, {Singh}, {Tominaga}, {Tanaka}, {Watson}, {Ahumada}, {Anupama}, {Ashall}, {Becerra}, {Bersier}, {Bhalerao}, {Bloom}, {Butler}, {Copperwheat}, {Coughlin}, {De}, {Drake}, {Duev}, {Frederick}, {Gonz{\'a}lez}, {Goobar}, {Heida}, {Ho}, {Horst}, {Hung}, {Itoh}, {Jencson}, {Kasliwal}, {Kawai}, {Khanam}, {Kulkarni}, {Kumar}, {Kumar}, {Kutyrev}, {Lee}, {Maeda}, {Mahabal}, {Murata}, {Neill}, {Ngeow}, {Penprase}, {Pian}, {Quimby}, {Ramirez-Ruiz}, {Richer}, {Rom{\'a}n-Z{\'u}{\~n}iga}, {Sahu}, {Srivastav}, {Socia}, {Sollerman}, {Tachibana}, {Taddia}, {Tinyanont}, {Troja}, {Ward}, {Wee}, \& {Yu}}]{Perley2019}
{Perley}, D.~A., {Mazzali}, P.~A., {Yan}, L., {et~al.} 2019, \mnras, 484, 1031, \dodoi{10.1093/mnras/sty3420}

\bibitem[{{Perley} {et~al.}(2021){Perley}, {Ho}, {Yao}, {Fremling}, {Anderson}, {Schulze}, {Kumar}, {Anupama}, {Barway}, {Bellm}, {Bhalerao}, {Chen}, {Duev}, {Galbany}, {Graham}, {Gromadzki}, {Guti{\'e}rrez}, {Ihanec}, {Inserra}, {Kasliwal}, {Kool}, {Kulkarni}, {Laher}, {Masci}, {Neill}, {Nicholl}, {Pursiainen}, {van Roestel}, {Sharma}, {Sollerman}, {Walters}, \& {Wiseman}}]{Perley2021}
{Perley}, D.~A., {Ho}, A. Y.~Q., {Yao}, Y., {et~al.} 2021, \mnras, \dodoi{10.1093/mnras/stab2785}

\bibitem[{{Piran} {et~al.}(2019){Piran}, {Nakar}, {Mazzali}, \& {Pian}}]{Piran2019}
{Piran}, T., {Nakar}, E., {Mazzali}, P., \& {Pian}, E. 2019, \apjl, 871, L25, \dodoi{10.3847/2041-8213/aaffce}

\bibitem[{{Piro} \& {Kollmeier}(2018)}]{Piro2018}
{Piro}, A.~L., \& {Kollmeier}, J.~A. 2018, \apj, 855, 103, \dodoi{10.3847/1538-4357/aaaab3}

\bibitem[{{Popham} {et~al.}(1999){Popham}, {Woosley}, \& {Fryer}}]{Popham1999}
{Popham}, R., {Woosley}, S.~E., \& {Fryer}, C. 1999, \apj, 518, 356, \dodoi{10.1086/307259}

\bibitem[{{Prentice} {et~al.}(2016){Prentice}, {Mazzali}, {Pian}, {Gal-Yam}, {Kulkarni}, {Rubin}, {Corsi}, {Fremling}, {Sollerman}, {Yaron}, {Arcavi}, {Zheng}, {Kasliwal}, {Filippenko}, {Cenko}, {Cao}, \& {Nugent}}]{Prentice2016}
{Prentice}, S.~J., {Mazzali}, P.~A., {Pian}, E., {et~al.} 2016, \mnras, 458, 2973, \dodoi{10.1093/mnras/stw299}

\bibitem[{{Prentice} {et~al.}(2018){Prentice}, {Maguire}, {Smartt}, {Magee}, {Schady}, {Sim}, {Chen}, {Clark}, {Colin}, {Fulton}, {McBrien}, {O'Neill}, {Smith}, {Ashall}, {Chambers}, {Denneau}, {Flewelling}, {Heinze}, {Holoien}, {Huber}, {Kochanek}, {Mazzali}, {Prieto}, {Rest}, {Shappee}, {Stalder}, {Stanek}, {Stritzinger}, {Thompson}, \& {Tonry}}]{Prentice2018}
{Prentice}, S.~J., {Maguire}, K., {Smartt}, S.~J., {et~al.} 2018, \apjl, 865, L3, \dodoi{10.3847/2041-8213/aadd90}

\bibitem[{{Ramirez-Ruiz} {et~al.}(2002){Ramirez-Ruiz}, {Celotti}, \& {Rees}}]{RamirezRuiz2002}
{Ramirez-Ruiz}, E., {Celotti}, A., \& {Rees}, M.~J. 2002, \mnras, 337, 1349, \dodoi{10.1046/j.1365-8711.2002.05995.x}

\bibitem[{{Rastinejad} {et~al.}(2025){Rastinejad}, {Levan}, {Jonker}, {Kilpatrick}, {Fryer}, {Sarin}, {Gompertz}, {Liu}, {Eyles-Ferris}, {Fong}, {Burns}, {Gillanders}, {Mandel}, {Malesani}, {O'Brien}, {Tanvir}, {Ackley}, {Aryan}, {Bauer}, {Bloemen}, {de Boer}, {Bom}, {Chac{\'o}n}, {Chambers}, {Chen}, {Chrimes}, {van Dalen}, {D'Elia}, {De Pasquale}, {Fulton}, {Groot}, {Gupta}, {Hartmann}, {van Hoof}, {Huber}, {Izzo}, {Jacobson-Galan}, {Jakobsson}, {Kong}, {Laskar}, {Lowe}, {Magnier}, {Maiorano}, {Martin-Carrillo}, {Mas-Ribas}, {Mata S{\'a}nchez}, {Nicholl}, {Nixon}, {Oates}, {Paek}, {Palmerio}, {Paris}, {Pieterse}, {Pugliese}, {Quirola Vasquez}, {van Roestel}, {Rossi}, {Rouco Escorial}, {Salvaterra}, {Schneider}, {Smartt}, {Smith}, {Smith}, {Srivastav}, {Torres}, {Ventura}, {Vreeswijk}, {Wainscoat}, {Yang}, \& {Yang}}]{Rastinejad2025}
{Rastinejad}, J.~C., {Levan}, A.~J., {Jonker}, P.~G., {et~al.} 2025, \apjl, 988, L13, \dodoi{10.3847/2041-8213/ade7f9}

\bibitem[{{Raynaud} {et~al.}(2020){Raynaud}, {Guilet}, {Janka}, \& {Gastine}}]{Raynaud2020}
{Raynaud}, R., {Guilet}, J., {Janka}, H.-T., \& {Gastine}, T. 2020, Science Advances, 6, eaay2732, \dodoi{10.1126/sciadv.aay2732}

\bibitem[{{Razzaque}(2010)}]{Razzaque2010}
{Razzaque}, S. 2010, \apjl, 724, L109, \dodoi{10.1088/2041-8205/724/1/L109}

\bibitem[{{Rees} \& {M{\'e}sz{\'a}ros}(1998)}]{Rees1998}
{Rees}, M.~J., \& {M{\'e}sz{\'a}ros}, P. 1998, \apjl, 496, L1, \dodoi{10.1086/311244}

\bibitem[{{Ricci} {et~al.}(2025){Ricci}, {Troja}, {Yang}, {Yadav}, {Liu}, {Sun}, {Wu}, {Gao}, {Zhang}, \& {Yuan}}]{Ricci2025}
{Ricci}, R., {Troja}, E., {Yang}, Y.-H., {et~al.} 2025, \apjl, 979, L28, \dodoi{10.3847/2041-8213/ad8b3f}

\bibitem[{{Ripperda} {et~al.}(2022){Ripperda}, {Liska}, {Chatterjee}, {Musoke}, {Philippov}, {Markoff}, {Tchekhovskoy}, \& {Younsi}}]{Ripperda2022}
{Ripperda}, B., {Liska}, M., {Chatterjee}, K., {et~al.} 2022, \apjl, 924, L32, \dodoi{10.3847/2041-8213/ac46a1}

\bibitem[{{Sari} \& {M{\'e}sz{\'a}ros}(2000)}]{Sari2000}
{Sari}, R., \& {M{\'e}sz{\'a}ros}, P. 2000, \apjl, 535, L33, \dodoi{10.1086/312689}

\bibitem[{{Sari} {et~al.}(1998){Sari}, {Piran}, \& {Narayan}}]{Sari1998}
{Sari}, R., {Piran}, T., \& {Narayan}, R. 1998, \apjl, 497, L17, \dodoi{10.1086/311269}

\bibitem[{{Schulze} {et~al.}(2011){Schulze}, {Klose}, {Bj{\"o}rnsson}, {Jakobsson}, {Kann}, {Rossi}, {Kr{\"u}hler}, {Greiner}, \& {Ferrero}}]{Schulze2011}
{Schulze}, S., {Klose}, S., {Bj{\"o}rnsson}, G., {et~al.} 2011, \aap, 526, A23, \dodoi{10.1051/0004-6361/201015581}

\bibitem[{{Shankar} {et~al.}(2025){Shankar}, {M{\"o}sta}, {Haas}, \& {Schnetter}}]{Shankar2025}
{Shankar}, S., {M{\"o}sta}, P., {Haas}, R., \& {Schnetter}, E. 2025, arXiv e-prints, arXiv:2504.11537, \dodoi{10.48550/arXiv.2504.11537}

\bibitem[{{Shibata} {et~al.}(2025){Shibata}, {Fujibayashi}, {Wanajo}, {Ioka}, {Lam}, \& {Sekiguchi}}]{Shibata2025}
{Shibata}, M., {Fujibayashi}, S., {Wanajo}, S., {et~al.} 2025, \prd, 111, 123017, \dodoi{10.1103/msy2-fwhx}

\bibitem[{{Shu} {et~al.}(2025){Shu}, {Yang}, {Yang}, {Xu}, {Chen}, {Eyles-Ferris}, {Dai}, {Yu}, {Shen}, {Sun}, {Ding}, {Zheng}, {Jiang}, {Li}, {Sun}, {Xu}, {Zhang}, {Jin}, {Rau}, {Wang}, {Wu}, {Yuan}, {Zhang}, {Nandra}, {Filippenko}, {Poidevin}, {Soria}, {Kumar}, {Aguado}, {An}, {An}, {An}, {Andrews}, {Anutarawiramkul}, {Baldini}, {Brink}, {Butpan}, {Cai}, {Castro-Tirado}, {Cheng}, {Cui}, {Farah}, {Fu}, {Fynbo}, {Gao}, {Han}, {Han}, {Howell}, {Hu}, {Jiang}, {Kumar}, {Lei}, {Li}, {Li}, {Liu}, {Liu}, {Liu}, {Liu}, {L{\'o}pez-Oramas}, {L{\'o}pez Fern{\'a}ndez-Nespral}, {Maund}, {McCully}, {Niu}, {Newsome}, {O'Brien}, {Pan}, {Pan}, {Padilla Gonzalez}, {P{\'e}rez-Fournon}, {Silima}, {Sun}, {Sun}, {Sun}, {Terreran}, {Tinyanont}, {Wang}, {Wang}, {Wang}, {Wiersema}, {Xu}, {Xue}, {Yang}, {Zhang}, {Zhang}, {Zhang}, {Zhang}, {Zhang}, {Zhao}, {Zhu}, {Xin}, {Yao}, {Cordier}, {Wei}, {Qiu}, \& {Daigne}}]{Xinwen2025}
{Shu}, X., {Yang}, L., {Yang}, H., {et~al.} 2025, \apjl, 990, L29, \dodoi{10.3847/2041-8213/adf4cd}

\bibitem[{{Srinivasaragavan} {et~al.}(2025){Srinivasaragavan}, {Hamidani}, {Schroeder}, {Sarin}, {Ho}, {Piro}, {Cenko}, {Anand}, {Sollerman}, {Perley}, {Maeda}, {O'Connor}, {Kuncarayakti}, {Miller}, {Ahumada}, {Vail}, {Duffell}, {Ghosh Dastidar}, {Andreoni}, {Bochenek}, {Brennan}, {Carney}, {Chen}, {Freeburn}, {Gal-Yam}, {Jacobson-Gal{\'a}n}, {Kasliwal}, {Li}, {Li}, {Sravan}, \& {Warshofsky}}]{Srinivasaragavan2025}
{Srinivasaragavan}, G.~P., {Hamidani}, H., {Schroeder}, G., {et~al.} 2025, arXiv e-prints, arXiv:2504.17516, \dodoi{10.48550/arXiv.2504.17516}

\bibitem[{{Srivastav} {et~al.}(2025){Srivastav}, {Chen}, {Gillanders}, {Rhodes}, {Smartt}, {Huber}, {Aryan}, {Yang}, {Beri}, {Cooper}, {Nicholl}, {Smith}, {Stevance}, {Carotenuto}, {Chambers}, {Aamer}, {Angus}, {Fulton}, {Moore}, {Smith}, {Young}, {de Boer}, {Gao}, {Lin}, {Lowe}, {Magnier}, {Minguez}, {Pan}, \& {Wainscoat}}]{Srivastav2025}
{Srivastav}, S., {Chen}, T.~W., {Gillanders}, J.~H., {et~al.} 2025, \apjl, 978, L21, \dodoi{10.3847/2041-8213/ad9c75}

\bibitem[{{Sun} {et~al.}(2025){Sun}, {Li}, {Liu}, {Gao}, {Wang}, {Yuan}, {Zhang}, {Filippenko}, {Xu}, {An}, {Ai}, {Brink}, {Liu}, {Liu}, {Wang}, {Wu}, {Wu}, {Yang}, {Zhang}, {Zheng}, {Ahumada}, {Dai}, {Delaunay}, {Elias-Rosa}, {Benetti}, {Fu}, {Howell}, {Huang}, {Kasliwal}, {Karambelkar}, {Stein}, {Lei}, {Lian}, {Peng}, {Frederiks}, {Ridnaia}, {Svinkin}, {Wang}, {Wang}, {Wei}, {An}, {Andrews}, {Bai}, {Dai}, {Ehgamberdiev}, {Fan}, {Farah}, {Feng}, {Fynbo}, {Guo}, {Guo}, {Hu}, {Hu}, {Jiang}, {Jin}, {Li}, {Li}, {Li}, {Liang}, {Ling}, {Liu}, {Mao}, {McCully}, {Mirzaqulov}, {Newsome}, {Padilla Gonzalez}, {Pan}, {Terreran}, {Tinyanont}, {Wang}, {Wang}, {Wen}, {Xiang}, {Xue}, {Yang}, {Zhu}, {Cai}, {Castro-Tirado}, {Chen}, {Chen}, {Chen}, {Chen}, {Chen}, {Chen}, {Chen}, {Cheng}, {Cordier}, {Cui}, {Cui}, {Dai}, {Fan}, {Feng}, {Guan}, {Han}, {Hou}, {Hu}, {Huang}, {Huo}, {Jia}, {Jia}, {Jiang}, {Jin}, {Jin}, {Kuulkers}, {Li}, {Li}, {Li}, {Li}, {Li}, {Li}, {Li}, {Liu}, {Liu}, {Liu}, {Liu}, {Lu}, {Luo}, {Ma}, {Mao},
  {Nandra}, {O'Brien}, {Pan}, {Rau}, {Rea}, {Sanders}, {Song}, {Sun}, {Sun}, {Tan}, {Tang}, {Tao}, {Wang}, {Wang}, {Wang}, {Wang}, {Wang}, {Wang}, {Xiong}, {Xu}, {Xu}, {Xu}, {Xu}, {Xu}, {Xue}, {Xue}, {Yan}, {Yang}, {Yang}, {Yang}, {Zhang}, {Zhang}, {Zhang}, {Zhang}, {Zhang}, {Zhang}, {Zhang}, {Zhang}, {Zhang}, {Zhang}, {Zhao}, {Zhao}, {Zhao}, {Zhao}, {Zhou}, {Zhu}, {Zhu}, \& {Zou}}]{Sun2025}
{Sun}, H., {Li}, W.~X., {Liu}, L.~D., {et~al.} 2025, Nature Astronomy, \dodoi{10.1038/s41550-025-02571-1}

\bibitem[{{Sun} {et~al.}(2022){Sun}, {Maund}, {Crowther}, \& {Liu}}]{Sun2022}
{Sun}, N.-C., {Maund}, J.~R., {Crowther}, P.~A., \& {Liu}, L.-D. 2022, \mnras, 512, L66, \dodoi{10.1093/mnrasl/slac023}

\bibitem[{{Suwa} \& {Tominaga}(2015)}]{Suwa2015}
{Suwa}, Y., \& {Tominaga}, N. 2015, \mnras, 451, 282, \dodoi{10.1093/mnras/stv901}

\bibitem[{{Suzuki} {et~al.}(2024){Suzuki}, {Irwin}, \& {Maeda}}]{Suzuki2024}
{Suzuki}, A., {Irwin}, C.~M., \& {Maeda}, K. 2024, \pasj, 76, 863, \dodoi{10.1093/pasj/psae055}

\bibitem[{{Suzuki} \& {Maeda}(2022)}]{Suzuki2022}
{Suzuki}, A., \& {Maeda}, K. 2022, \apj, 925, 148, \dodoi{10.3847/1538-4357/ac3d8d}

\bibitem[{{Svinkin} {et~al.}(2024){Svinkin}, {Frederiks}, {Lysenko}, {Ridnaia}, {Tsvetkova}, {Ulanov}, {Cline}, \& {Konus-Wind Team}}]{Svinkin2024}
{Svinkin}, D., {Frederiks}, D., {Lysenko}, A., {et~al.} 2024, GRB Coordinates Network, 35972, 1

\bibitem[{{Symbalisty}(1984)}]{Symbalisty1984}
{Symbalisty}, E.~M.~D. 1984, \apj, 285, 729, \dodoi{10.1086/162551}

\bibitem[{{Taddia} {et~al.}(2015){Taddia}, {Sollerman}, {Leloudas}, {Stritzinger}, {Valenti}, {Galbany}, {Kessler}, {Schneider}, \& {Wheeler}}]{Taddia2015}
{Taddia}, F., {Sollerman}, J., {Leloudas}, G., {et~al.} 2015, \aap, 574, A60, \dodoi{10.1051/0004-6361/201423915}

\bibitem[{{Taddia} {et~al.}(2019){Taddia}, {Sollerman}, {Fremling}, {Barbarino}, {Karamehmetoglu}, {Arcavi}, {Cenko}, {Filippenko}, {Gal-Yam}, {Hiramatsu}, {Hosseinzadeh}, {Howell}, {Kulkarni}, {Laher}, {Lunnan}, {Masci}, {Nugent}, {Nyholm}, {Perley}, {Quimby}, \& {Silverman}}]{Taddia2019}
{Taddia}, F., {Sollerman}, J., {Fremling}, C., {et~al.} 2019, \aap, 621, A71, \dodoi{10.1051/0004-6361/201834429}

\bibitem[{{Tchekhovskoy} \& {McKinney}(2012)}]{Tchekhovskoy2012}
{Tchekhovskoy}, A., \& {McKinney}, J.~C. 2012, \mnras, 423, L55, \dodoi{10.1111/j.1745-3933.2012.01256.x}

\bibitem[{{Tchekhovskoy} {et~al.}(2011){Tchekhovskoy}, {Narayan}, \& {McKinney}}]{Tchekhovskoy2011}
{Tchekhovskoy}, A., {Narayan}, R., \& {McKinney}, J.~C. 2011, MNRAS, 418, L79, \dodoi{10.1111/j.1745-3933.2011.01147.x}

\bibitem[{{Uzdensky} \& {MacFadyen}(2007)}]{Uzdensky2007}
{Uzdensky}, D.~A., \& {MacFadyen}, A.~I. 2007, \apj, 669, 546, \dodoi{10.1086/521322}

\bibitem[{{van Dalen} {et~al.}(2025){van Dalen}, {Levan}, {Jonker}, {Malesani}, {Izzo}, {Sarin}, {Quirola-V{\'a}squez}, {Mata S{\'a}nchez}, {de Ugarte Postigo}, {van Hoof}, {Torres}, {Schulze}, {Littlefair}, {Chrimes}, {Ravasio}, {Bauer}, {Martin-Carrillo}, {Fraser}, {van der Horst}, {Jakobsson}, {O'Brien}, {De Pasquale}, {Pugliese}, {Sollerman}, {Tanvir}, {Zafar}, {Anderson}, {Galbany}, {Gal-Yam}, {Gromadzki}, {M{\"u}ller-Bravo}, {Ragosta}, \& {Terwel}}]{vanDalen2025}
{van Dalen}, J. N.~D., {Levan}, A.~J., {Jonker}, P.~G., {et~al.} 2025, \apjl, 982, L47, \dodoi{10.3847/2041-8213/adbc7e}

\bibitem[{{Wanderman} \& {Piran}(2010)}]{Wanderman2010}
{Wanderman}, D., \& {Piran}, T. 2010, \mnras, 406, 1944, \dodoi{10.1111/j.1365-2966.2010.16787.x}

\bibitem[{{Wang} {et~al.}(2016){Wang}, {Han}, {Xu}, {Wang}, {Dai}, {Wu}, \& {Wei}}]{Wang2016}
{Wang}, L.-J., {Han}, Y.-H., {Xu}, D., {et~al.} 2016, \apj, 831, 41, \dodoi{10.3847/0004-637X/831/1/41}

\bibitem[{{Wheeler} {et~al.}(2000){Wheeler}, {Yi}, {H{\"o}flich}, \& {Wang}}]{Wheeler2000}
{Wheeler}, J.~C., {Yi}, I., {H{\"o}flich}, P., \& {Wang}, L. 2000, \apj, 537, 810, \dodoi{10.1086/309055}

\bibitem[{{White} {et~al.}(2022){White}, {Burrows}, {Coleman}, \& {Vartanyan}}]{White2022}
{White}, C.~J., {Burrows}, A., {Coleman}, M. S.~B., \& {Vartanyan}, D. 2022, \apj, 926, 111, \dodoi{10.3847/1538-4357/ac4507}

\bibitem[{{Woosley}(1993)}]{Woosley1993}
{Woosley}, S.~E. 1993, \apj, 405, 273, \dodoi{10.1086/172359}

\bibitem[{{Woosley} \& {Bloom}(2006)}]{Woosley2006}
{Woosley}, S.~E., \& {Bloom}, J.~S. 2006, \araa, 44, 507, \dodoi{10.1146/annurev.astro.43.072103.150558}

\bibitem[{{Wu} {et~al.}(2025){Wu}, {Yu}, {Liu}, {Dai}, {Lei}, {Wu}, {Xu}, {Zhang}, {Zhu}, \& {Zou}}]{Wu2025}
{Wu}, G.-L., {Yu}, Y.-W., {Liu}, L.-D., {et~al.} 2025, arXiv e-prints, arXiv:2505.12491.
\newblock \doarXiv{2505.12491}

\bibitem[{{Yadav} {et~al.}(2025){Yadav}, {Troja}, {Ricci}, {Yang}, {Veres}, {Wieringa}, {O'Connor}, {Kang}, {Becerra}, {Ryan}, \& {Busmann}}]{Yadav2025}
{Yadav}, M., {Troja}, E., {Ricci}, R., {et~al.} 2025, arXiv e-prints, arXiv:2505.08781, \dodoi{10.48550/arXiv.2505.08781}

\bibitem[{{Yao} {et~al.}(2022){Yao}, {Ho}, {Medvedev}, {Nayana}, {Perley}, {Kulkarni}, {Chandra}, {Sazonov}, {Gilfanov}, {Khorunzhev}, {Khatami}, \& {Sunyaev}}]{Yao2021}
{Yao}, Y., {Ho}, A. Y.~Q., {Medvedev}, P., {et~al.} 2022, \apj, 934, 104, \dodoi{10.3847/1538-4357/ac7a41}

\bibitem[{{Yin} {et~al.}(2024){Yin}, {Zhang}, {Yang}, {Sun}, {Zhang}, {Shao}, {Hu}, {Zhu}, {Xu}, {An}, {Gao}, {Wu}, {Zhang}, {Castro-Tirado}, {Pandey}, {Rau}, {Lei}, {Xie}, {Ghirlanda}, {Piro}, {O'Brien}, {Troja}, {Jonker}, {Yu}, {An}, {Chen}, {Chen}, {Dong}, {Eyles-Ferris}, {Fan}, {Fu}, {Fynbo}, {Gao}, {Huang}, {Jiang}, {Jiang}, {Julakanti}, {Kuulkers}, {Lao}, {Li}, {Ling}, {Liu}, {Liu}, {Mou}, {Pan}, {Wei}, {Wu}, {Yadav}, {Yang}, {Yuan}, \& {Zhang}}]{Yin2024}
{Yin}, Y.-H.~I., {Zhang}, B.-B., {Yang}, J., {et~al.} 2024, \apjl, 975, L27, \dodoi{10.3847/2041-8213/ad8652}

\bibitem[{{Zenati} {et~al.}(2020){Zenati}, {Siegel}, {Metzger}, \& {Perets}}]{Zenati2020}
{Zenati}, Y., {Siegel}, D.~M., {Metzger}, B.~D., \& {Perets}, H.~B. 2020, \mnras, 499, 4097, \dodoi{10.1093/mnras/staa3002}

\bibitem[{{Zheng} {et~al.}(2025){Zheng}, {Zhu}, {Lu}, \& {Zhang}}]{Zheng2025}
{Zheng}, J.-H., {Zhu}, J.-P., {Lu}, W., \& {Zhang}, B. 2025, arXiv e-prints, arXiv:2503.24266, \dodoi{10.48550/arXiv.2503.24266}

\bibitem[{{Zhu} {et~al.}(2025){Zhu}, {Zheng}, \& {Zhang}}]{Zhu2025}
{Zhu}, J.-P., {Zheng}, J.-H., \& {Zhang}, B. 2025, arXiv e-prints, arXiv:2507.18544.
\newblock \doarXiv{2507.18544}

\end{thebibliography}

\appendix
\section{Collapsar jet propagation in circumstellar medium}\label{sec:pluto}

The scenario in which a relativistic jet is choked during its propagation through a dense CSM has been proposed as a potential explanation for certain optical counterparts of EP transients. In particular, \citet{Srinivasaragavan2025} suggested that the choked jet scenario could account for the optical emission observed in EP250108a, provided that a CSM with mass $ M_m = 0.07\,\msun $ extending out to $ R_m = 3.7\times 10^{13}\,\cm $ is sufficient to choke a jet with energy $ E_{\rm jet} = 5.7\times 10^{51}\,{\rm erg} $. Similarly, \citet{Hamidani2025} proposed that a CSM of $ M_m = 0.03\,\msun $ extending to $ R_m = 3\times 10^{13}\,\cm $ can marginally choke a jet with a two-sided luminosity $ L = 2\times 10^{50}\,\erg\,\s^{-1} $, opening angle $ \theta_0 = 0.17\,{\rm rad} $ and duration of $ T = 100\,\s $. These parameters are consistent with analytic estimates of jet suffocated within the CSM \citep{Nakar2015, Hamidani2025a}. However, using numerical simulations of jet propagation through dense media, \citet{Duffell2020} have found that choking typically requires extreme CSM masses, well above those inferred from observations. On the other hand, their simulations were conducted under axisymmetric conditions, which may introduce numerical artifacts affecting jet propagation. Axisymmetry suppresses mixing between the jet and the surrounding material, allowing the jet to remain more stable and propagate faster. At the same time, it artificially concentrates mass along the polar axis, thereby slowing down the jet propagation \citep[see][for a detailed discussion]{Gottlieb2021a}. Therefore, assessing the conditions under which jet choking occurs may require fully three-dimensional simulations. More recently, \citet{Suzuki2022, Suzuki2024} performed high-resolution 3D simulations of jet propagation through the CSM. For the relevant scenario considered above, they found that the jet decelerates to $\Gamma \approx 30$, consistent with the findings of \citet{Hamidani2025}. However, their jet was launched with a relatively low maximum Lorentz factor of $\Gamma = 100$, suggesting that a jet with an initial Lorentz factor of several hundred may still emerge from the CSM with $\Gamma \gtrsim 100$, sufficient to power a GRB.

\subsection{Numerical Setup}
We perform a 3D relativistic hydrodynamic simulation of a relativistic jet breaking out from a dense star into a surrounding CSM. Our initial setup is inspired by the configurations explored in \citet{Hamidani2025} and \citet{Srinivasaragavan2025}, with the key modification of increasing the CSM mass by an order of magnitude to promote jet choking, as proposed in those studies. The jet is injected into a progenitor star with radius $ R_\star = 10^{11}\,\cm $ and total mass of  $ M_\star = 12\,\msun $, characterized by a density profile $ \rho \sim r^2(1-r/R_\star)^3 $. The jet is launched conically within an opening angle $ \theta_0 = 0.15\,{\rm rad} $ from an inner boundary at $ z_0 = 10^9\,{\rm cm} $, and is active for $ T = 100\,{\rm s} $. It has a two-sided luminosity $ L = 2\times 10^{50}\,{\rm erg\,s^{-1}} $, initial Lorentz factor of $ \Gamma_0 = 5 $, and specific enthalpy of $ h_0 = 100 $, corresponding to an asymptotic Lorentz factor $ \Gamma_\infty = \Gamma_0 h_0 = 500 $. The total injected energy is thus $ E = LT = 10^{52}\,{\rm erg} $. After breaking out of the star, the jet propagates into a dense CSM with total mass $ M_m = 1\,\msun $, extending to $ R_m = 3.7\times 10^{13}\,\cm $ and following a density profile $ \rho \sim r^{-2} $.

We use a 3D Cartesian grid implemented in the public code PLUTO v4.0 \citep{Mignone2007}, employing an ideal equation of state with an adiabatic index of 4/3, the Harten-Lax-van Leer (HLL) Riemann solver, piecewise parabolic interpolation, and third-order Runge-Kutta time integration. The computational grid is divided into three patches along each spatial axis. Along the $ x $ and $ y $ directions, the innermost region extends to $ \pm 2\times 10^{9}\,\cm $, discretized with 100 uniformly spaced cells. Beyond this, the outer regions span from $ \lvert 2\times 10^{9}\,\cm\rvert $ to $ \lvert 3\times 10^{13}\,\cm\rvert $, each resolved with 400 logarithmically spaced cells. Along the $ z $-axis, the first patch extends from the injection point $ z_0 $ to $ 5\,z_0 $ with 100 uniformly spaced cells. This is followed by a second uniform patch consisting of 500 cells extending up to the stellar surface at $ z = R_\star $. The final patch stretches logarithmically to the outer CSM boundary at $ z = R_m $ with 500 cells. The jet is injected at $ z_0 $ with a cylindrical injection radius of $ r_0 = z_0\theta_0 $. We caution that our jet may be underresolved at large radii, and a dedicated study is needed to validate these numerical results.

\subsection{Results}
We find that the CSM decelerates the jet head to mildly relativistic velocities during its propagation (the jet bulk behind the head is ultra-relativistic). However, the jet head is moving faster than what is assumed in analytic estimates. This is at least partly due to the jet accelerating as it propagates through the star, resulting in a narrower post-breakout opening angle upon entering the CSM, $ \theta_b \approx \theta_0/3 $ \citep[see also][]{Gottlieb2019b}, which has not been taken into account in previous analytic studies. For a wind-like CSM density profile, $ \rho = \rho_0 r^{-2} $, where $ \rho_0 = M_m/4\pi R_m $, the Lorentz factor of a \emph{relativistic} collimated jet head follows \citep{Harrison2018}
\begin{equation}\label{eq:Gammah}
    \Gamma_h \approx 0.83\left(\frac{4\pi R_m L}{M_m c^3\theta_b^4}\right)^{1/10}\approx 1.66\left(\frac{R_mL}{M_m c^3\theta_0^4}\right)^{1/10}\,,
\end{equation}
demonstrating that the correction of $ \theta \approx \theta_b \approx \theta_0/3 $ translates to a correction of almost two orders of magnitude in the required CSM mass or jet luminosity to choke the jet.
The criterion for a relativistic jet breakout from the CSM is
\begin{equation}\label{eq:Tbo}
    T - t_b \gtrsim \frac{R_m}{2\Gamma_h^2 c} \approx 0.18 \left(\frac{M_m R_m^4\theta_0^4}{Lc^2}\right)^{1/5}\,, 
\end{equation}
where $ t_b $ is the breakout time from the dense star (before entering the CSM), during which the jet head is subrelativistic. Our numerical results are consistent with Equation~\eqref{eq:Gammah}, yielding $ \Gamma_h \approx 3 $. This implies that the deceleration is insufficient for the jet tail to catch up and cross the reverse shock at the jet head, which is the condition for jet choking.

\begin{figure}[]
  \centering
  \includegraphics[width=0.48\textwidth]{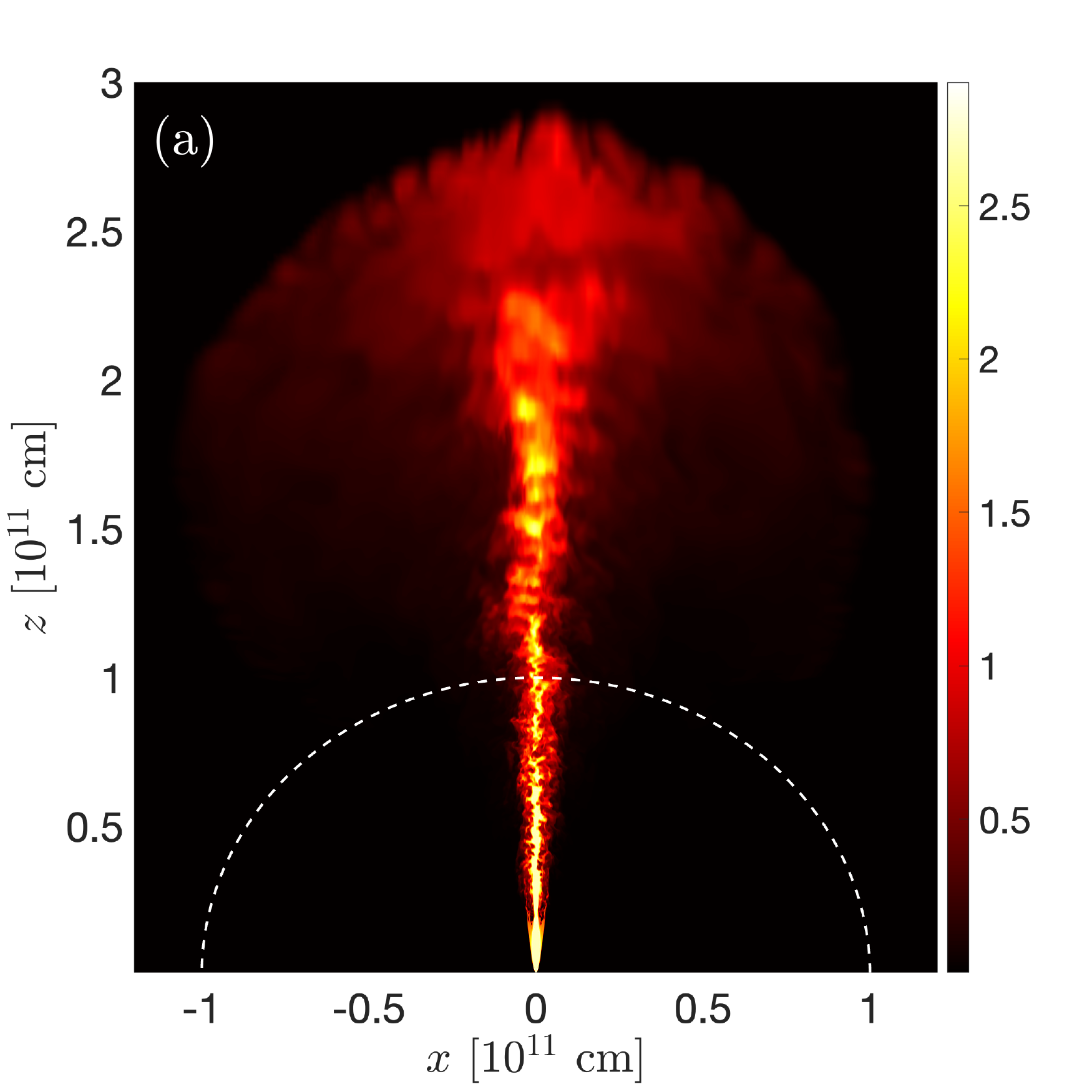}
  \includegraphics[width=0.48\textwidth]{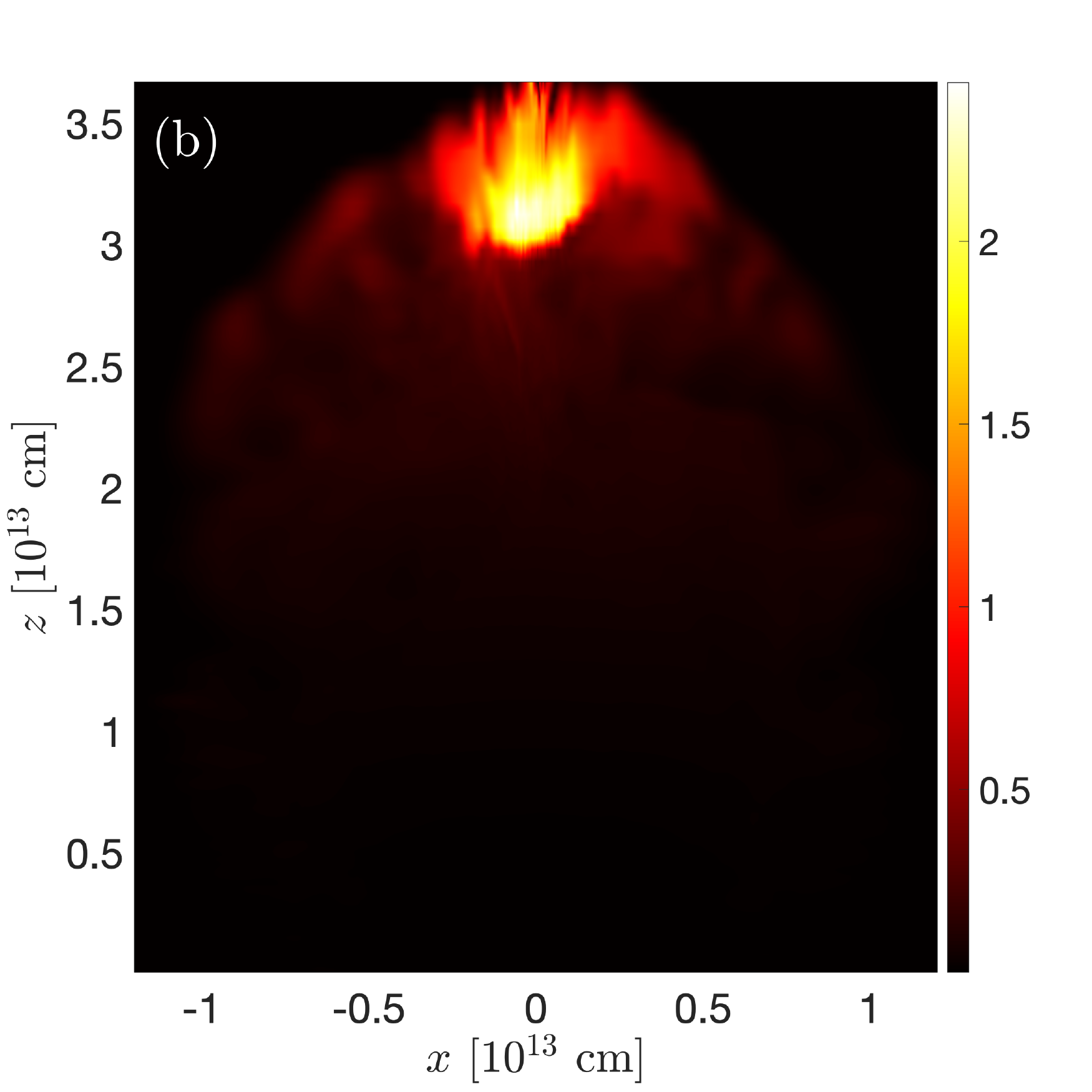}
  \caption{Meridional maps of $ {\rm log}(\Gamma_\infty) = {\rm log}(h\Gamma)$  {(\bf a)} shortly after the jet breaks out of the star (dashed white line) at $ t = 6\,\s $ after launching; and  {\bf (b)} upon reaching the outer edge of the CSM at $ t = 20\,{\rm minutes} $. While the jet head is mildly relativistic, the bulk of the jet retains ultra-relativistic velocities throughout its interaction with the CSM, indicating that the surrounding medium is insufficiently massive to choke the jet.}
  \label{fig:RHD}
\end{figure}

Figure~\ref{fig:RHD} displays maps of the asymptotic Lorentz factor of the outflow, from shortly after it escapes the dense stellar envelope (panel a) to $ \sim 2.5 $ orders of magnitude later, as it reaches the outer edge of the CSM (panel b). The persistence of ultra-relativistic velocities in both panels indicates that, even with a CSM an order of magnitude more massive than those considered in \citet{Hamidani2025, Srinivasaragavan2025}, the surrounding medium remains incapable of choking the relativistic jet.

Due to the computational cost of such large-scale 3D simulations, we have only performed a single simulation tailored to the scenario relevant to the EP transient under consideration. Nevertheless, our results align with the axisymmetric simulations of \citet{Duffell2020}, who showed that analytic models in the \emph{Newtonian} jet head regime overestimate the choking ability of the CSM. Their results suggest that, in order to choke the jet, the CSM mass or radius must be increased, or the jet luminosity decreased, by a factor of $ \sim 25 $ compared to the analytic prediction in \citet{Nakar2015}, roughly consistent with the required correction of $ \theta \approx \theta_0/3 $. This leads to a revised choking condition for a Newtonian jet head, as proposed in \citet{Duffell2020},
\begin{equation}
    T \lesssim 30\,\left(\frac{M_m}{10^{-2}\,\msun}\frac{R_m}{10^{13.5}\,\cm}\frac{10^{51}\,\erg\,\s^{-1}}{L_{\rm iso}}\right)^{1/2}\,\s,
\end{equation}
where $ L_{\rm iso} $ is the isotropic equivalent luminosity of the jet. This result implies that typical CSM environments are unlikely to choke jets with characteristic GRB properties. Consequently, it places constraints on the occurrence rate of jets choked by CSM and challenges their role as a dominant mechanism for powering electromagnetic transients.

\end{document}